%% file: main.tex
\documentclass[conference]{IEEEtran}
\IEEEoverridecommandlockouts
\usepackage{cite}
\usepackage{amsmath,amssymb,amsfonts}
\usepackage{algorithmic}
\usepackage{graphicx}
\usepackage{textcomp}
\usepackage{xcolor}
\def\BibTeX{{\rm B\kern-.05em{\sc i\kern-.025em b}\kern-.08em
    T\kern-.1667em\lower.7ex\hbox{E}\kern-.125emX}}

\usepackage{hyperref}
\usepackage{cleveref}
\usepackage{xcolor}
\usepackage{algorithm}
\usepackage{algorithmic}
\usepackage{footnote}
\usepackage{subfigure}

\begin{document}

\bstctlcite{IEEEexample:BSTcontrol}

\title{Fine-Grained Scheduling for Containerized HPC Workloads in Kubernetes Clusters\\
}

\author{\IEEEauthorblockN{Peini Liu\IEEEauthorrefmark{1}\IEEEauthorrefmark{2},
Jordi Guitart\IEEEauthorrefmark{1}\IEEEauthorrefmark{2},
\\
}
\IEEEauthorblockA{\IEEEauthorrefmark{1}Barcelona Supercomputing Center, Barcelona, Spain}
\IEEEauthorblockA{\IEEEauthorrefmark{2}Universitat Polit\`ecnica de Catalunya, Barcelona, Spain}

E-mail: \{peini.liu, jordi.guitart\}@bsc.es}

\maketitle

\begin{abstract}
Containerization technology offers lightweight OS-level virtualization, and enables portability, reproducibility, and flexibility by packing applications with low performance overhead and low effort to maintain and scale them. Moreover, container orchestrators (e.g., Kubernetes) are widely used in the Cloud to manage large clusters running many containerized applications. However, scheduling policies that consider the performance nuances of containerized High Performance Computing (HPC) workloads have not been well-explored yet.
This paper conducts fine-grained scheduling policies for containerized HPC workloads in Kubernetes clusters, focusing especially on partitioning each job into a suitable multi-container deployment according to the application profile. We implement our scheduling schemes on different layers of management (application and infrastructure), so that each component has its own focus and algorithms but still collaborates with others.
Our results show that our fine-grained scheduling policies 
outperform baseline and baseline with CPU/memory affinity enabled policies, reducing the overall response time by 35\% and 19\%, respectively, and also improving the makespan by 34\% and 11\%, respectively. They also provide better usability and flexibility to specify HPC workloads than other comparable HPC Cloud frameworks, while providing better scheduling efficiency thanks to their multi-layered approach.
\end{abstract}


\input{1.introduction}
\input{2.background}

\input{3.architecture}

\input{4.implementation}
\input{5.evaluation}

\input{7.conclusion}

\section*{Acknowledgment}
We thank Lenovo for providing the technical infrastructure to run the experiments in this paper. This work was partially supported by Lenovo as part of Lenovo-BSC collaboration agreement, by the Spanish Government under contract PID2019-107255GB-C22, and by the Generalitat de Catalunya under contract 2017-SGR-1414 and under grant 2020 FI-B 00257.

\bibliographystyle{IEEEtran}
\bibliography{IEEEabrv,bib}

\end{document}

%% file: 1.introduction.tex
\section{Introduction}\label{sec:introduction}

Modern computing infrastructure is evolving at a fast pace to Cloud computing services. Containerization, as a fundamental technology for Cloud computing, allows efficient utilization and easy maintenance of the infrastructure. So far, this attractive paradigm has also had an impact on High Performance Computing (HPC)\cite{Asch2018}\cite{Data2019}.

Previous works have demonstrated the possibility to enable HPC workloads on Cloud infrastructure using containers\cite{beltre2019}, and have discussed some best practices for HPC workloads on the Cloud\cite{bestpractice2022}\cite{hpccloud2016}. 
The deployment of containerized HPC workloads in the Cloud is done by container orchestrators, which have the capability to launch and manage containers and their full life cycles, and leverage resource availability and the user specifications to decide the placement of containers. Several orchestrators are available nowadays such as Docker Swarm\cite{dockerswarm}, Mesos\cite{mesos1}, and Kubernetes\cite{k8s}. Kubernetes has been widely adopted in commercial production systems
, such as Google Kubernetes Engine\cite{gke}, 
and provides a wide and active toolkit ecosystem.

Currently, Kubernetes is not optimized for the management of HPC applications, as it was designed to support the autonomous management of loosely-coupled long-lived online microservices, enabling their self-healing and auto-scaling. Although it also includes some support for short-lived batch jobs, the tuning of their specification, scheduling, and management must rely on other algorithms and tools. For example, Kubeflow MPI operator\cite{kubeflowmpi} provides a specification for MPI applications through the execution a standalone worker and Volcano\cite{volcanompi} provides some plugins to enable optimized scheduling for jobs. As the HPC community has important performance considerations on its workloads, developing new deployment schemes for different types of HPC workloads that improve their performance is needed.

Our previous systematical performance studies \cite{Liu2020}\cite{Liu2021} have demonstrated through standalone executions that some types of containerized HPC applications achieve better performance when exploiting multi-container deployments which partition the processes that belong to each application into multiple containers in each node and when constraining each of those containers to a single NUMA (Non-Uniform Memory Access) domain or pinning them to specific processors. However, these deployment schemes have not yet been integrated in multiprogrammed environments for HPC workloads by current Cloud orchestrators.

In this paper, we look for fine-grained scheduling policies for allocating containerized HPC workloads through Kubernetes. The goal is to introduce our optimized management framework to inspire HPC community developers and operators on how to deploy their workloads in a fine-grained way to improve performance and leverage containerization and orchestration technologies. Our main contributions are:
\begin{itemize}
    \item We present a two-layer scheduling architecture. In the application layer, an agent decides the wrapping granularity of the HPC workload based on the characteristics of the applications. In the infrastructure layer, an MPI-aware plugin and task-group scheduling scheme are enabled within a containerized platform scheduler. The MPI-aware plugin decides each MPI task-container mapping and the resource requirements/limits of each container. The task-group scheduling scheme is used to allocate containers to available nodes.
    \item We establish a real platform (so-called Scanflow(MPI)-Kubernetes), implement the algorithms in both layers, and evaluate our fine-grained scheduling scheme for containerized HPC workloads deployments.
\end{itemize}

%% file: 2.background.tex
\section{Background and Related Work}\label{sec:background}

\subsection{Orchestration of Containers in Kubernetes}
Containerization is a lightweight virtualization technology that builds upon resource isolation and limitation features of the Linux kernel, such as \verb|namespaces| and \verb|cgroups|, respectively. Currently, containerization is widely used to pack applications because of its portability, isolation, and high availability. Generally, there are two types of containerized applications running in the Cloud.

\begin{itemize}
    \item Long-lived online microservices: loose-coupled services, each of them being an independent module to be deployed or managed in the long term (e.g., Web services).
    \item Short-lived batch jobs: batched processing jobs, each of them comprising a batch of tasks that are executed once and then terminate (e.g., MapReduce, MPI, and Spark).
\end{itemize}

Kubernetes 
supports both types of applications. From the users' perspective, they submit their specifications of services or jobs to Kubernetes, which is responsible for encapsulating them into containers that are wrapped in Pods to be deployed in the nodes. From the providers' perspective, all the nodes and resources are controlled by Kubernetes. Whenever there is a request, Kubernetes managers have to generate the Pod specification for each type of job or service, and select the node (using a scheduling policy to filter and rank nodes) to run each Pod, so that the Kubernetes node agent (i.e., Kubelet) can launch the Pod in the selected node.

\subsection{Enabling HPC Workloads in Kubernetes}
HPC workloads are considered as batch jobs in Kubernetes. An HPC workload is specified as a launcher and one or multiple workers. Each launcher or worker is a container that can be executed as a Pod and run in parallel in a Kubernetes cluster. 
However, the original Kubernetes batch jobs are not designed for supporting the HPC applications efficiently. The specification for HPC applications is limited, thereby relevant application-related information cannot be considered while scheduling. Also, the Kubernetes default scheduler does not schedule jobs but individual Pods. Thus, some add-ons have been designed by the community to enhance the usability when specifying and allocating HPC workloads.

Kubeflow MPI operator\cite{kubeflowmpi} provides a better specification for MPI jobs which defines an MPI 'Launcher' and an MPI 'Worker'. In most cases, all the MPI worker processes will be launched in this 'Worker' container. Moreover, Kubeflow MPI operator mounts the ssh folder for all Pods belonging to the job through a Kubernetes Secret to establish the communication. But this operator does not enhance the Kubernetes default scheduler, thus the allocation of Pods is not considered at the MPI job level but for each individual Pod.

Volcano \cite{volcanompi} is an add-on for running HPC workloads on Kubernetes. It features batch scheduling capabilities (such as gang scheduling to make sure that a job will start to run only when all its tasks are ready to be deployed) that Kubernetes scheduler does not support, and also integrates some Big Data/AI frameworks in its controller. 
Moreover, it provides ssh/service plugins to deal with the Pods' connection and permissions and with the service discovery, and features a customizable scheduler, so that the system operator can choose different strategies for job scheduling. 

\subsection{Deployment and Scheduling Schemes for Containerized HPC Workloads}

Former works in this area have focused on deploying containerized HPC workloads in traditional HPC systems. These systems have batch-oriented workload managers or resource managers, such as Slurm \cite{slurm} or Torque, and some of them have included container support \cite{slurmcontainer}. The convergence between HPC systems and Cloud environments has been also explored \cite{Zhou2021}\cite{dividenodes}, but these works mainly divide the nodes into clusters for different usage and enable the access to the HPC cluster from the Cloud cluster. 

Once the HPC workloads are containerized, they could run on Cloud environments by directly using container orchestration platforms.
Beltre et al. \cite{Data2019}\cite{pankaj2018} did some performance analysis on enabling HPC workloads on Cloud infrastructure. They analyzed the HPC workload performance while using different container orchestrators like Kubernetes and Docker Swarm and different networks like InfiniBand. They used the Kubernetes default scheduler. 
Misale et al. \cite{kubeflux} introduced KubeFlux, a Kubernetes plugin scheduler that is based on Flux graph-based scheduler. This plugin translates the Pod into a Flux job and uses the policy within Fluxion to allocate jobs.
Saha et al. \cite{Saha2019} showed how MPI applications can be scheduled by Mesos using a policy-based approach. 

There are also some works focusing on the policies for scheduling HPC jobs in the Cloud. For instance,
Gupta et al.\cite{hpccloud2016} presented novel heuristics for online application-aware job scheduling in multi-platform environments.
Fu et al. \cite{yuqi2019} proposed a progress-based container placement for short-lived containerized jobs. 
Aupy et al.\cite{researve2019} provided an optimal job reservation strategy in scheduling to minimize the cost.


\begin{figure*}[hbtp]
    \centering
    \includegraphics[width=0.8\textwidth]{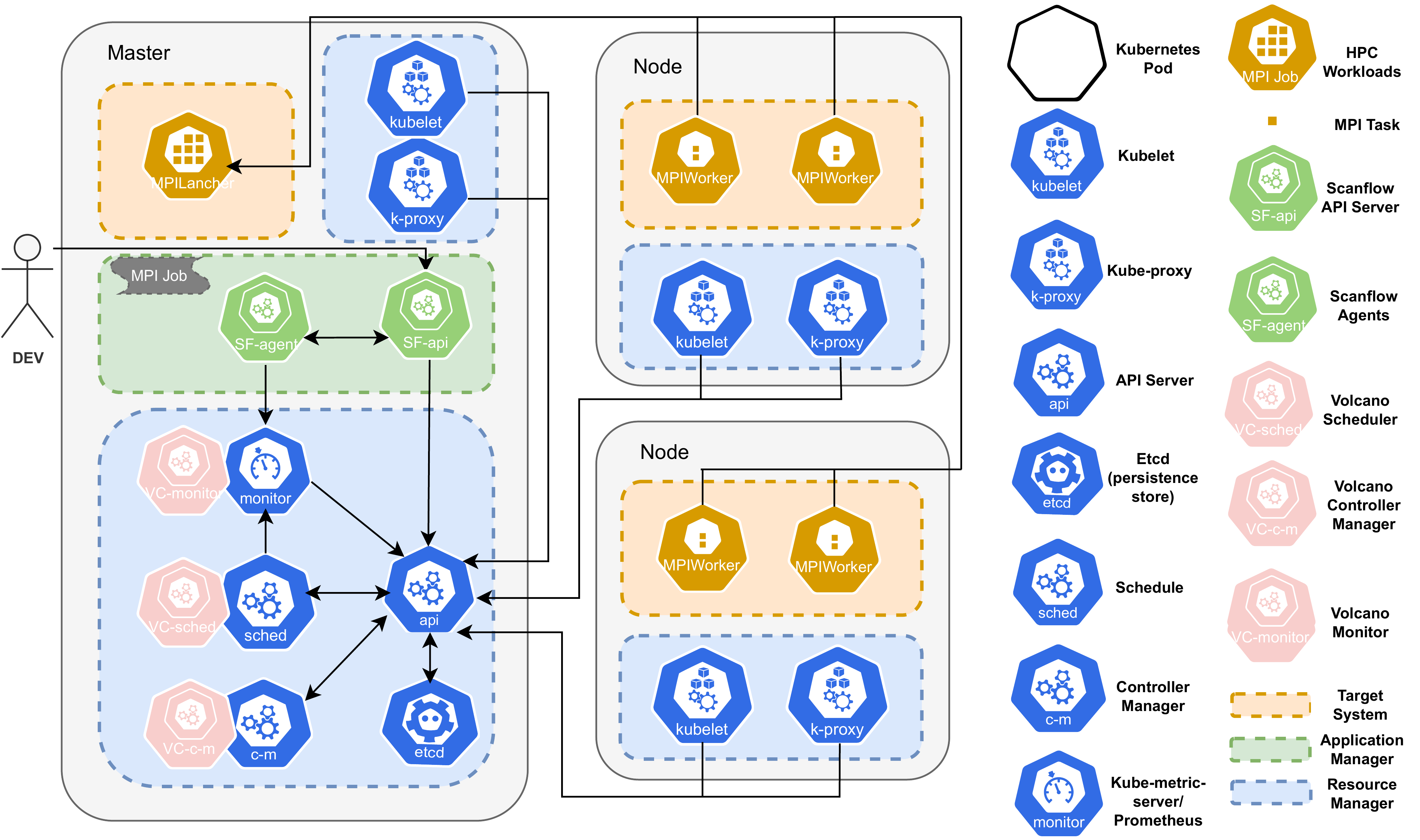}
    \caption{Scanflow(MPI)-Kubernetes: A practical platform for managing HPC workloads.}
    \label{fig:architecture}
\end{figure*}

HPC community has important performance considerations on its workloads. Therefore, trialing new deployment schemes for different types of HPC workloads to improve their performance is necessary. Walkup et al. \cite{bestpractice2022} reported best practices for running compute-, memory-, and network-intensive HPC workloads on the Cloud. Medel et al. \cite{MEDEL2018286} conducted a performance analysis over Kubernetes considering the deployment and initialization overhead as well as understanding the performance of different Pod settings. Moreover, they provided a rule to decide the number of containers per pod by considering the characteristics of the application. 
Our previous papers\cite{Liu2020}\cite{Liu2021} demonstrated through standalone executions that some types of containerized HPC applications achieve better performance when exploiting multi-container deployments which partition the processes that belong to each application into multiple containers in each node, and when constraining each of those containers to a single NUMA domain or pinning them to specific processors. These works show some ways to achieve better performance for HPC workloads in the Cloud, but those insights have not yet been integrated and utilized by the current Cloud orchestrators.

%% file: 3.architecture.tex
\section{System Architecture}\label{sec:archi}
\label{III}

Our fine-grained scheduling approach for containerized HPC workloads is built over the existing Scanflow-Kubernetes platform\cite{scanflowdemopaper}\cite{liuccgrid}. It is implemented both within a Scanflow(MPI) extension package in the application layer (see in Scanflow-Kubernetes github repository\footnote{https://github.com/bsc-scanflow/scanflow/tree/mpi}) and an enhanced Volcano scheduler/controller manager in the infrastructure layer (see in Volcano github repository\footnote{https://github.com/peiniliu/volcano/tree/peini}).
The whole architecture of this platform is depicted in Fig. \ref{fig:architecture}.


\textbf{Target System:} The yellow area in the figure shows the target system focusing on the HPC workloads. 
From a static design perspective, HPC workloads are defined as distributed jobs. Typically, an MPI job in the Cloud is composed of a launcher and one or several workers, and all the MPI processes of the job are executed within the workers\cite{kubeflow}. However, following the idea of using containerized instances to decouple the processes and considering the potential benefits of multi-container deployments for HPC workloads\cite{Liu2020}\cite{Liu2021}, each worker can be split into several finer-grained workers which hold part of processes and are executed in parallel on each node. 
From a dynamic implementation perspective, the launcher and workers of a job are conducted as containerized instances (i.e., Kubernetes Pods) executing together in the Cloud. 
All the Pods belonging to the job run once for each time the job is submitted.

\textbf{Application Manager:} Application manager (i.e., Scanflow) is used as a controller of the application layer, as shown in the green area of Fig. \ref{fig:architecture}. 
To work with HPC workloads, we implemented a Scanflow(MPI) extension\footnotemark[1], which allows the users to use the Scanflow-client Python library to easily define and build HPC workloads locally and submit MPI jobs to Scanflow-server to be deployed. This server can connect with Scanflow-agents to calculate proper MPI job granularity (number of workers and nodes to be used) and also submit jobs to Kubernetes Control Plane to run them in a Kubernetes cluster. We also added support for HPC workloads in Scanflow through a granularity-aware planner agent, which can decide the proper granularity for each user-submitted MPI job by considering the provided application profile and the status of the cluster nodes (see Algorithm \ref{algo:1} in Section \ref{sec:scheduling algorithm}).

\textbf{Resource Manager:} The blue area of Fig. \ref{fig:architecture} shows the resource manager (i.e., Kubernetes) on the infrastructure layer. Thanks to the Scanflow(MPI) extension described above, our HPC workloads are well-wrapped into containers, thus we can directly use a container orchestrator (i.e., Kubernetes) as resource manager to finely manage the job deployment. We can also take advantage from the wide range of toolkits in the Kubernetes ecosystem, such as Volcano and Prometheus\footnote{https://prometheus.io/}. 


Kubernetes Control Plane manages the cluster and responds to cluster events. By default, it includes the API Server, etcd database, and more relevant to this work, Controller Manager and Scheduler. Each type of object has its own Controller Manager to watch its life-cycle, for example, Volcano job controller manager watches the job object, and create the pods to run master/workers to completion. Scheduler watches pods without node assigned and selects the best node for each pod to run on. Node selection has two steps: filtering (to find a set of nodes that are feasible to place the pod) and scoring (to rank the nodes to choose the most suitable placement).


Kubernetes is originally used to manage microservices, thus the default controllers (i.e., Deployment, ReplicaSet) are intended to manage and scale these applications. Similarly, the default scheduler policies are also well-fitted for deploying this type of long-running microservices. However, the usability of the default Job controller and scheduler for deploying HPC workloads is limited. To cope with this, we leverage Volcano into our platform to evolve default jobs into Volcano jobs and change the default scheduler into Volcano scheduler. 
We also take advantage of the Volcano feature to support additional scheduling plugins to implement an MPI-aware plugin inside the Volcano job controller to configure the \textit{hostfile} and resource request of each worker. Additionally, a task-group scheduling plugin is also implemented inside Volcano scheduler to make scalable and balanced scheduling for fine-grained Volcano jobs (see Section \ref{sec:scheduling algorithm}). 

After the global scheduling decided by the Kubernetes Control Plane, pods are stored inside etcd indicating their assigned node. The next step is to start the pod in the corresponding node through the Kubelet component, which
is used for maintaining the pods on nodes (e.g., starting, terminating, reporting). By default, the pods could use requested resources from the whole single node (but not more than their specified limit). However, to do a finer-grain deployment, a CPU/memory management policy should be configured.


HPC workloads can move to different CPUs and increase the context switches if using shared resources, which will degrade the workload performance. As well, related work has shown that CPU/memory affinity could help HPC workloads to gain performance\cite{Liu2020}\cite{Liu2021}. Consequently, we explore different Kubelet settings to allocate exclusive CPUs or using NUMA affinity. 
This paper evaluates two Kubelet settings: (1) default: all pods could use shared resources in a node under the resource limits specification; (2) CPU/memory affinity: sets \verb|--cpu-manager-policy=static| and \verb|--topology-manager-policy=best-effort|, so that a pod will be allocated on exclusive CPUs and try best-effort to use CPUs from a single NUMA node.

%% file: 4.implementation.tex
\section{Fine-Grained Scheduling}\label{sec:scheduling algorithm}

\begin{figure}
    \centering
    \includegraphics[width=0.9\linewidth]{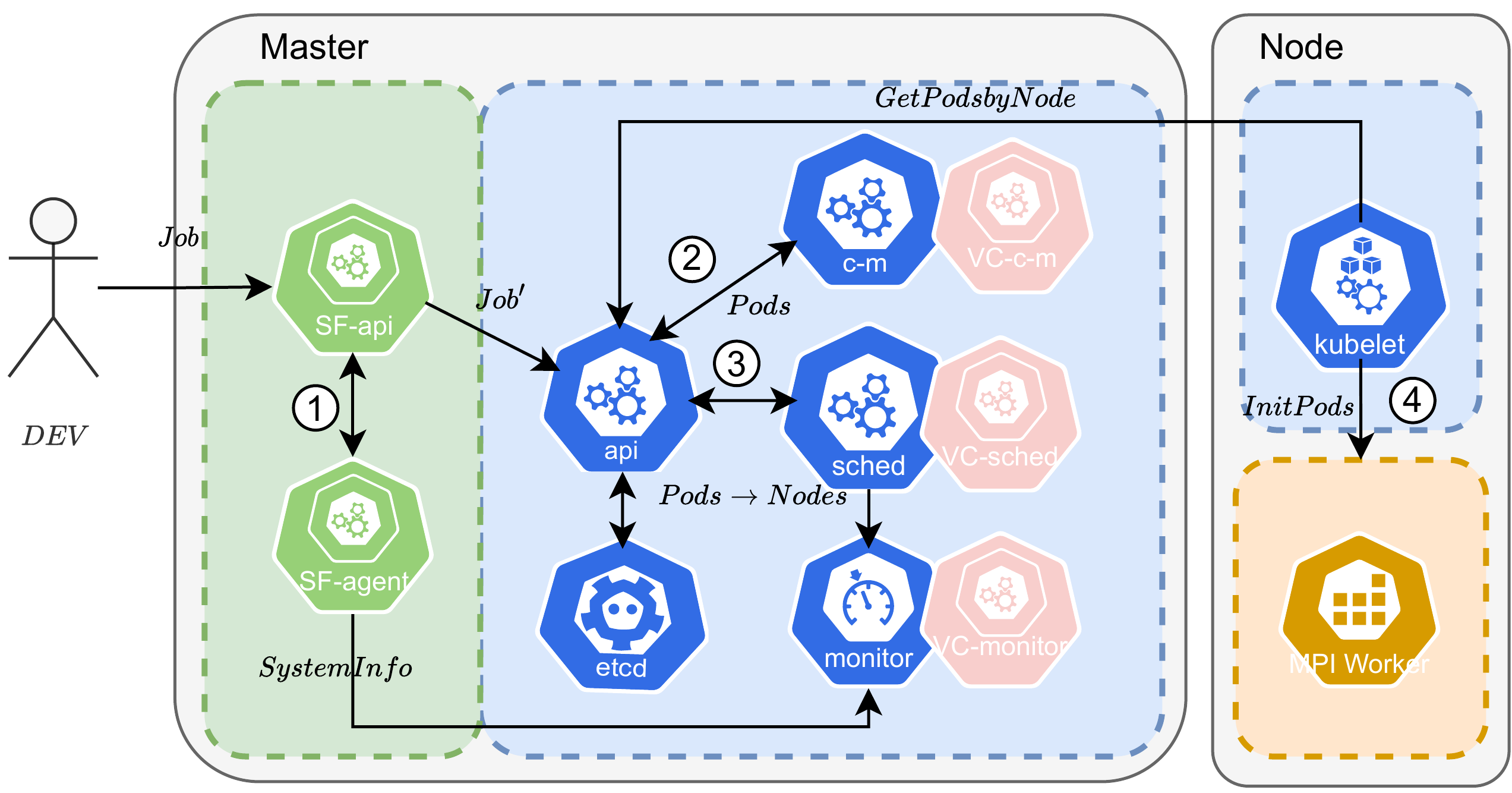}
    \caption{Scheduling steps for HPC workloads deployment.}
    \label{fig:algorithm}
\end{figure}

Fine-grained scheduling for containerized HPC workloads is composed of several steps which are shown in Fig. \ref{fig:algorithm}. The notations used are explained in Table \ref{tab:notation}. In the master, the application manager and the resource manager components use their global views to decide the nodes where to allocate the pods belonging to the job, while in each node, the resource manager will decide the resources actually used for each pod that is allocated on. 

\begin{table}[htbp]
\caption{Notation table. }
\centering
\begin{tabular}{p{2.9cm}|p{4.9cm}}
\hline
\textbf{Notation} & \textbf{Explanation} \\ \hline
$Job$                &  MPI Job metadata.                   \\
$N_{t}$      &  Number of tasks for the Job (fixed).             \\
$N_{n}$  &  Number of nodes for the Job.\\
$N_{w}$  & Number of workers for the Job.\\
$N_{g}$  & Number of groups of Pods for the Job.\\
$R(cpu, memory)$     & Resource requirements/limits for the Job.   \\
$Pods$ & Units to wrap master/workers of the Job. \\
$Pod_{w}^{i}$ & Worker $i$ of the Job. \\
$Pods_{w}$ & Workers of the Job. \\
$Pod_{l}$ & Launcher of the Job. \\
$Node_{j}$ & Node $j$ in the cluster. \\
$Nodes$ & Nodes in the cluster. \\
Map($Pod_{w}^{i}$ $\to$ $Node_{j}$) & Mapping of worker $i$ allocated to a node $j$. \\ 
\hline
\end{tabular}
\label{tab:notation}
\end{table}

\subsection{Application layer granularity selection algorithm}


In the application layer, the developer defines the MPI job (i.e., $Job$), including $N_t$, which is fixed as it specifies the number of MPI processes this application will start (same as calling ‘mpirun -np 16’), and the profile of the application (e.g., network, CPU, memory intensive), which implicitly defines the relevant QoS, and submits it to the Scanflow API Server. 
The Scanflow(MPI) planner agent is responsible for the automatic calculation of other parameters related with the construct/granularity of the $Job$ according to a predefined policy set by the admin (see step 1 in Fig. \ref{fig:algorithm}), as described in Algorithm \ref{algo:1}. In particular, it calculates $N_w$, $N_g$, and $N_n$. For that purpose, the planner agent considers $N_t$, the application profile, and its resource requirements. If desired, the user can provide a default value for $N_w$ and the agent can get the maximum $N_n$ from Prometheus. 

We define two policies, "scale" and "granularity", to determine $N_w$. In both, each network-intensive application will be packed into a single worker, while the CPU-intensive and the memory-intensive applications will be split into multiple workers, with $N_w=N_n$ in the "scale" policy, and $N_w=N_t$ in the "granularity" policy. If no policy is set, the agent will keep the default $N_w$ specified by the user. 
Finally, the updated MPI job with granularity will be submitted to the Scanflow API Server, which will transmit $Job$ to a Kubernetes cluster through the Kubernetes Control Plane.

\begin{algorithm}
\caption{Granularity Selection (\textit{Planner} agent)}
\begin{algorithmic}[1]
\renewcommand{\algorithmicrequire}{\textbf{Input:}}
\renewcommand{\algorithmicensure}{\textbf{Output:}}
\REQUIRE $Job$: MPI Job metadata, $SystemInfo$: System information,
        $Policy$: Granularity policy, $Profile$: Job profile \\
\ENSURE  $Job$: Updated MPI Job metadata with granularity\\
\textit{\COMMENT{\% Agent Sensor: get job specs and system information}}\\
\STATE 
  $N_t$, $N_w$ $\gets$ $Job$\;   
\STATE
  $N_n$ $\gets$ $SystemInfo$\; 
\\
\textit{\COMMENT{\% Agent Rule: set granularity according to job profile}}\\
\IF {($Policy =$ "scale")}
    \IF {($Profile =$ "network")}
    \STATE $N_n = 1$, $N_w = 1$, $N_g = 1$
    \ELSIF{($Profile =$ "CPU" $||$ "memory")}
    \STATE $N_n = min(N_n,N_t)$, $N_w = N_n$, $N_g=N_n$
    \ENDIF
\ELSIF{($Policy =$ "granularity")}
    \IF {($Profile =$ "network")}
    \STATE $N_n = 1$, $N_w = 1$, $N_g = 1$
    \ELSIF{($Profile =$ "CPU" $||$ "memory")}
    \STATE $N_n = min(N_n,N_t)$, $N_w = N_t$, $N_g=N_n$
    \ENDIF
\ELSE
\STATE $N_n = 1$, $N_w = N_w$, $N_g=N_n$
\ENDIF
\\ \COMMENT{\textit{\% Agent Actuator: update and submit the job}}\\
\STATE $Job$ $\gets$ Update($N_n$, $N_w$, $N_g$)   
\STATE Submit($Job$) to Scanflow API Server 
\end{algorithmic} 
\label{algo:1}
\end{algorithm}

\subsection{Infrastructure layer task-group scheduling}

In the infrastructure layer, Kubernetes with enhanced Volcano is used to control the life-cycle of the $Job$ (see step 2 in Fig. \ref{fig:algorithm}) and decide the best nodes to place the $Job$ (see step 3 in Fig. \ref{fig:algorithm}). Volcano job controller manager watches the $Job$ and creates a $Pod$ for each MPI launcher/worker within the job. However, $Job$ needs some dynamic configuration while generating the $Pods$. Thus, we enhanced Volcano job controller manager with a plugin implementing Algorithm \ref{algo:2} to make it MPI-aware. This plugin helps $Job$ to allocate $N_t$ into $N_w$ in a RoundRobin fashion, decide the $R(cpu, memory)$ for each worker, as well as generate the \textit{hostfile} for all the workers to communicate. After the initialization of the $Job$, all its launcher/workers are wrapped as $Pods$ that are registered in Kubernetes API Server and wait for Volcano scheduler to choose the allocated node.

\begin{algorithm}
\caption{Dynamic MPI-aware Job Controller}
\begin{algorithmic}[1]
\renewcommand{\algorithmicrequire}{\textbf{Input:}}
\renewcommand{\algorithmicensure}{\textbf{Output:}}
\REQUIRE $Job$: Job metadata with granularity\\
\ENSURE  $Pods$: Updated pods with resources, $Hostfile$: Hostfile for MPI application to allocate tasks\\
\textit{\COMMENT{\% Step 1: get job specification}}\\
\STATE 
   $Pod_{l}$, $Pods_{w}$, $N_t$, $N_w$, $N_n$, $R(cpu/N_t, memory/N_t)$ $\gets$ $Job$\;    
\\ \textit{\COMMENT{\% Step 2: allocate tasks into workers in RoundRobin}}
\STATE $nTasksInWorker$ $\gets$ AllocateTasks($N_t$, $N_w$)  
\\ \textit{\COMMENT{\% Step 3: set up pod resources and the hostfile according to the number of tasks allocated }}
\FOR {$i$ in $0$ to $N_w-1$}
\STATE $nTasks$ $\gets$ GetnTasks($nTasksInWorker$, $i$)
\STATE $Pod_{w}^{i}$ $\gets$ Update($Pod_{w}^{i}$, $R(cpu/N_t \cdot nTasks, memory/N_t \cdot nTasks)$) 
\STATE $Hostfile$ $\gets$ Add(Hostname($Pod_{w}^{i}$), slots=$nTasks$)
\ENDFOR
\STATE $Pods$ = $Pods_{w}$ + $Pod_{l}$
\RETURN $Pods$ 
\end{algorithmic} 
\label{algo:2}
\end{algorithm}

Pods are the smallest deployable entities in Kubernetes, so the scheduler decides their placement individually. However, when enabling granularity, there are various pods that belong to the same job, and we also aim to scale evenly the job into multiple nodes. Therefore, we implemented a task-group plugin inside Volcano (see Algorithm \ref{algo:3}). The idea is to group evenly the workers into multiple groups, enabling node affinity for the workers within each group and node anti-affinity among groups. This is done in two steps. First, building multiple groups for every job and allocating worker pods into those groups. Then, filtering for each pod the nodes where it is feasible to schedule it (using Kubernetes default filter), scoring those nodes (using the procedure described in Algorithm \ref{algo:4}), and selecting the best one.

\begin{algorithm}
\caption{Task-Group Scheduling}
\begin{algorithmic}[1]
\renewcommand{\algorithmicrequire}{\textbf{Input:}}
\renewcommand{\algorithmicensure}{\textbf{Output:}}
\REQUIRE $N_g$: Number of groups, $Pods_{w}$: Worker pods, $Nodes$: Nodes\\
\ENSURE  $Pods_{w}$: Worker pods with nodes assigned allocated
\\ \textit{\COMMENT{\% Step 1: build and allocate workers into groups}}
\STATE $groups$ $\gets$ newGroups($N\_g$)  
\FOR {$i$ in $Pods_{w}$} 
\STATE $groups$ $\gets$ sortGroupByResourceRequests($groups$) 
\STATE $selected\_group$ = $groups[0]$ 
\STATE AddWorkerToGroup($Pod_{w}^{i}$, $selected\_group$)
\ENDFOR
\\ \textit{\COMMENT{\% Step 2: predicate and priority node for worker}}
\STATE $Pods_{w}$ $\gets$ WorkerOrderFn($groups$) 
\FOR {$i$ in $Pods_{w}$}
\FOR {$j$ in $Nodes$}
\STATE $pre\_nodes$ $\gets$ PredicateFn($Pod_{w}^{i}$, $Node_j$) 
\ENDFOR
\FOR {$k$ in $pre\_nodes$}
\STATE $node\_score$ $\gets$ NodeOrderFn($Pod_{w}^{i}$, $pre\_nodes_k$) 
\ENDFOR
\STATE $best\_node$ $\gets$ getBestNode($max(node\_score)$) 
\STATE $Pod_{w}^{i}$ $\gets$ Update(Map($Pod_{w}^{i}$, $best\_node$))
\ENDFOR
\RETURN $Pods_{w}$ 
\end{algorithmic} 
\label{algo:3}
\end{algorithm}

\begin{algorithm}
\caption{NodeOrderFn Node Score Calculation}
\begin{algorithmic}[1]
\renewcommand{\algorithmicrequire}{\textbf{Input:}}
\renewcommand{\algorithmicensure}{\textbf{Output:}}
\REQUIRE $worker$: Worker, $node$: Node\\
\ENSURE  $score$: score of worker allocated to node \\
\STATE $group$ $\gets$ getGroupByWorker($worker$)  
\\ \textit{\COMMENT{\% Step 1: base score is the number of bound task in the same group that allocated in the node}}
\STATE $bound\_nodes$ $\gets$ getNodesBoundbyGroup($group$) 
\FOR{$bound\_node$ in $bound\_nodes$}
\IF{$bound\_node$ = $node$}
\STATE $score ++$ 
\ENDIF
\ENDFOR
\\ \textit{\COMMENT{\% Step 2: count remaining tasks in the same group}}
\STATE  $score = score +$ len($group$.$worker$) 
\\ \textit{\COMMENT{\% Step 3: avoid other groups in the node}}
\FOR{ $allocated\_group$ in getGroupsInNode($node$) }
\IF{$allocated\_group \ne group$}
\STATE $score --$ 
\ENDIF
\ENDFOR
\RETURN $score$
\end{algorithmic} 
\label{algo:4}
\end{algorithm}

Algorithm \ref{algo:3} and Algorithm \ref{algo:4} call some auxiliary functions: ‘sortGroupByResourceRequests’ sorts the groups from big to small according to their resource request so that the workers can be evenly added to the groups and each group has similar resource requests; ‘WorkerOrderFn’ decides the order of the workers taking into account that they can belong to different groups, so it picks up a group and enqueues the workers within the group instead of ordering the workers just by using its id; ‘PredicateFn’ filters the nodes available to allocate some pods by constraints such as node taints or tolerations; ‘NodeOrderFn’ in Algorithm \ref{algo:4} calls ‘getNodesBoundbyGroup’, which returns the node that has been already assigned to the pods in the group, so that when deciding the next pod in the group, the bound node has a higher score.

\subsection{Node affinity settings}

In each node, Kubelet takes a set of $Pods$ that are provided through the API Server, and starts the containers described in those pods (see step 4 in Fig. \ref{fig:algorithm}). By default, the containers could use requested resources from the whole single node (but not more than the limit), but this paper considers different Kubelet settings to allocate exclusive CPUs or using NUMA affinity for containers, as introduced in Section \ref{sec:archi}. 

%% file: 5.evaluation.tex
\section{Evaluation}\label{sec:evaluation}

In this section, we evaluate the performance of our proposed fine-grained scheduling policies for containerized HPC workloads through typical HPC MPI benchmarks.

\subsection{Experimental Platform}

\textbf{Hardware:} Our experiments are executed on a five-node K8s cluster. Each host consists of 2 x Intel 2697v4 CPUs (18 cores each, hyperthreading disabled), 256 GB RAM, 60 TB GPFS file system, and 1-Gigabit Ethernet network.

\textbf{Software:} All the nodes run CentOS release 7.7.1908 with host kernel 3.10.0-1062.el7.x86\_64. The Scanflow(MPI)-Kubernetes platform is built based on Kubernetes v1.19.16 (with Docker 19.03.11, Etcd 3.4.9, Flannel 0.15.0, CNI 0.8.6, and CoreDNS 1.7.0). Its corresponding toolkits (as described in Section \ref{III}) are Prometheus v14.3.0 and our enhancement of Volcano\footnotemark[2] based on v1.5.0. Additionally, we use Scanflow(MPI)\footnotemark[1] version with built-in planner agent.



\subsection{Experimental Setup and Metrics}
\textbf{Kubernetes Cluster Settings}: Our Kubernetes cluster comprises five nodes. We dedicate one node to hold the Control Plane and execute the launcher of MPI applications while the other four nodes are used to run the workers of MPI applications. For each node, we reserve four cores for system and Kubernetes components, thus 32 cores (16 from each socket) can be used for the allocation of MPI workloads.

As described in Section \ref{III}, by default Kubelet sets CPU/memory affinity as none. For those experiments that require enabling CPU/memory affinity inside Kubelet, we configure it as \verb|--cpu-manager-policy=static| and \verb|--topology-manager-policy=best-effort|.

\textbf{Scheduler Settings}: 
We use Volcano as default scheduler in the baseline experiments. Volcano is configured by default with the gang plugin enabled, whereas the allocations of all the workers remain the same as Kubernetes default scheduler.


Our fine-grained scheduling policies use two-level scheduling. In the application layer, the granularity selection algorithm is implemented inside the Scanflow planner agent. In the infrastructure layer, we use an enhanced version of Volcano that implements our MPI-aware controller and also features our task-group scheduling.

\textbf{Benchmark Settings}: We use the HPC Challenge benchmark suite\footnote{http://icl.cs.utk.edu/hpcc/} and the MiniFE proxy application for unstructured implicit finite element codes\footnote{https://github.com/Mantevo/miniFE}. They are built with OpenMPI v4.0.3rc3, and run with 16 MPI processes in exactly-subscribed mode, all of them bound to all the processors allocated to the application (i.e., 16 cores) in all the scenarios. 

\begin{figure}[htbp]
    \centering
    \includegraphics[width=0.75\linewidth]{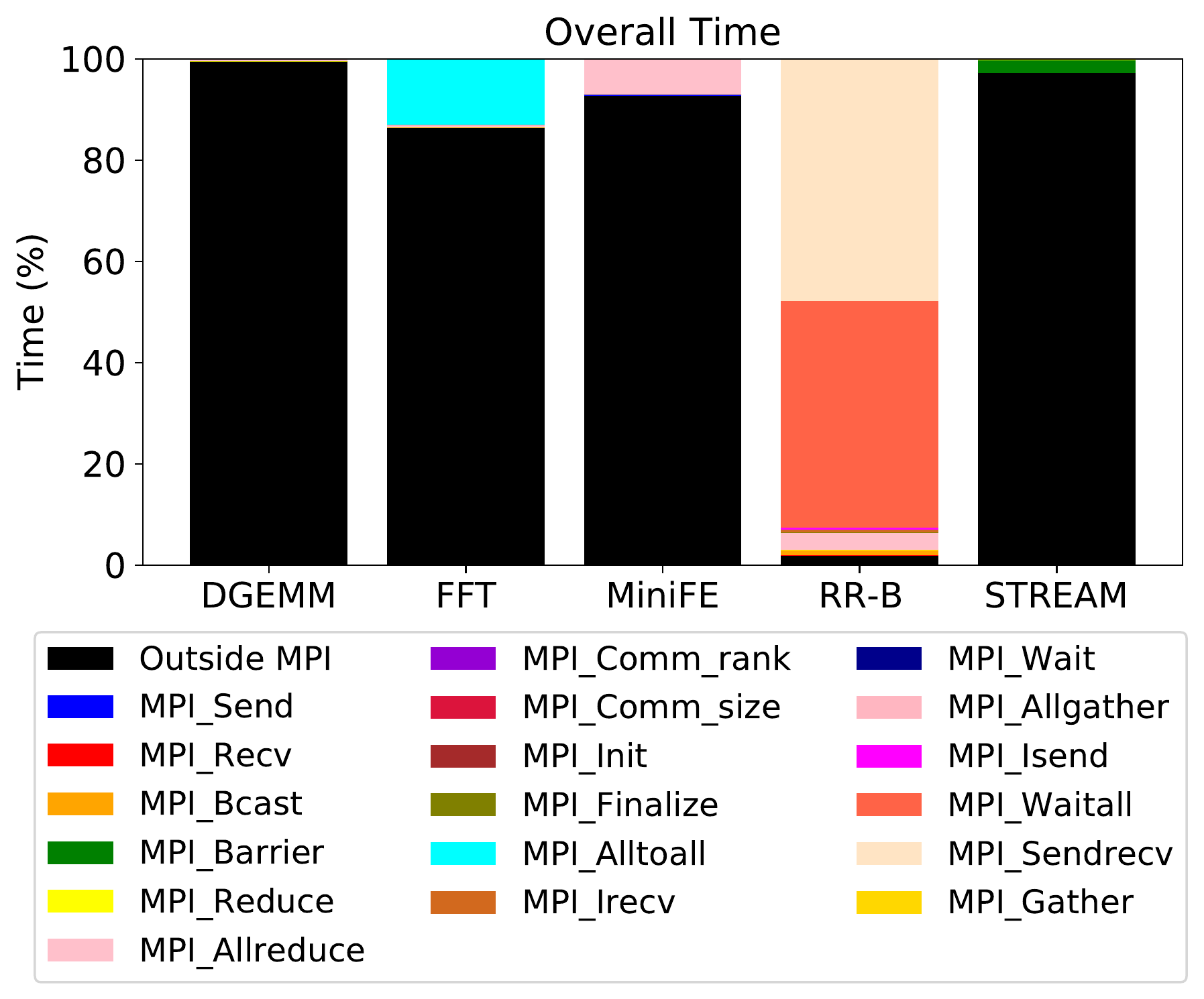}
    \caption{Benchmarks MPI Profiling Analysis.}
    \label{fig:performance}
\end{figure}

The specific MPI profile analysis (used to classify MPI applications) can be found in Fig. \ref{fig:performance} and our paper\cite{Liu2020}. EP-DGEMM and EP-STREAM are MPI throughput applications: the former is CPU intensive and the latter is memory bandwidth intensive. G-RandomRing Bandwidth and G-FFT are MPI communication applications where processes need to communicate (frequently and globally) with each other.
For application MiniFE, we set problem size as nx=ny=nz=512. As shown in Fig. \ref{fig:performance}, it contains some MPI\_Allreduce communications (i.e., global reduce) but they can scale without introducing much network latency\cite{allreduce}. Thus the application is categorized as memory and CPU intensive.

\textbf{Scenario Settings}:
We consider six scenarios (see Table \ref{tab:scenario}): \textit{NONE} and \textit{CM} are two baseline scenarios, the former with the default settings of Kubernetes, and the latter with the CPU/memory affinity settings supported by Kubelet. Given the well-known benefit of tuning the CPU/memory affinity for HPC MPI applications, we compare our policies on top of CPU/memory affinity. Scenarios \textit{CM\_S} and \textit{CM\_G} use two different strategies for agent granularity selection, namely 'scale'(S) and 'granularity'(G), which were described in Algorithm \ref{algo:1}. Scenarios \textit{CM\_S\_TG} and \textit{CM\_G\_TG} maintain the benefits that the granularity policies apply in the application layer and also show the effectiveness of using our proposed task-group scheduling (TG) (see Algorithms \ref{algo:3}-\ref{algo:4}) in the infrastructure layer. Summarizing, the settings of the six scenarios are as shown in Table \ref{tab:scenario}:

\begin{table}[htbp]
\caption{Scenarios settings. }
\centering
\begin{tabular}{p{1.3cm}|p{1.5cm}|p{2cm}|p{2.2cm}}
\hline
\textbf{Scenarios} & \textbf{Kubelet} & \textbf{Scanflow} & \textbf{Volcano} \\ \hline
NONE               & default & &             default(gang)     \\
CM                & cpu/memory affinity &   &     default(gang)            \\
CM\_S                 & cpu/memory affinity  & granularity selection 'scale'   & default(gang)                   \\
CM\_G                & cpu/memory affinity     &  granularity selection 'granularity'  & default(gang)                  \\
CM\_S\_TG                 & cpu/memory affinity   & granularity selection 'scale'   & default(gang)+task-group scheduling                  \\
CM\_G\_TG                & cpu/memory affinity    &  granularity selection 'granularity'  & default(gang)+task-group scheduling                 \\
\hline
\end{tabular}
\label{tab:scenario}
\end{table}

\textbf{Metrics}: We consider four main metrics in our evaluation:
\begin{itemize}
    \item \textbf{Job Running Time ($T_{i}^{r}$):} the running performance of job $i$.
    \item \textbf{Job Response Time ($T_{i}$):} the total wallclock time from the instant at which the job $i$ is submitted to the system until it terminates\cite{metrics}. It is composed of two parts: the time $T_{i}^{w}$ that job $i$ is waiting and the time $T_{i}^{r}$ that job $i$ is actually running in parallel on multi-processing nodes. Thus, $T_{i} = T_{i}^{w}+T_{i}^{r}$.
    \item \textbf{Overall Response Time ($T$):} the total response time summed from all the jobs. $T = \sum{T_{i}}$
    \item \textbf{Makespan ($T_{makespan}$)}: the time required for all jobs to terminate. It is directly linked with utilization and throughput, and each can be derived from the others\cite{metric2}. 
\end{itemize}


\subsection{Experiment 1: Schedule with one type of MPI workload}

Our previous paper \cite{Liu2020} showed that EP-DGEMM benchmark can improve its performance thanks to a finer-grain deployment scheme. Thus, firstly we set an experiment with this single type of application, and we submit 10 MPI EP-DGEMM jobs with an arrival interval of 60 seconds.

\begin{figure}[htbp]
    \centering
    \includegraphics[width=0.65\linewidth]{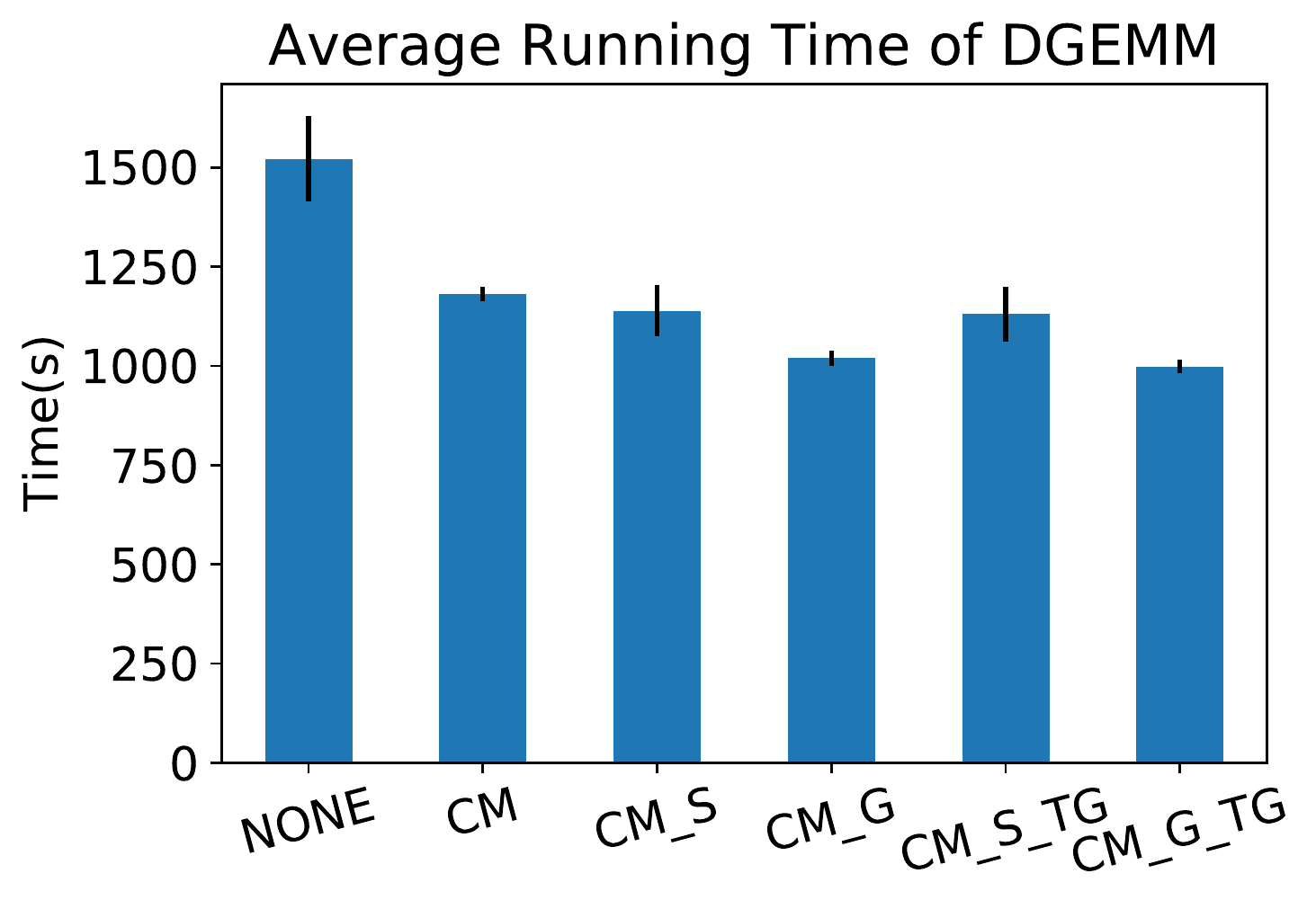}
    \caption{Average job running time of 10 EP-DGEMM jobs.}
    \label{fig:exp1-performance}
\end{figure}

Fig. \ref{fig:exp1-performance} shows the average performance of the EP-DGEMM workload in the six scenarios. Scenario \textit{CM} shows better cache utilization (less L3 misses), more local memory accesses, and less remote memory accesses than \textit{NONE} scenario. When enabling 'scale' and 'granularity' policies, we partition each application within a larger number of containers, but fewer number of processes per container. Those scenarios, namely \textit{CM\_S*} (i.e., \textit{CM\_S} and \textit{CM\_S\_TG}) and \textit{CM\_G*} (i.e., \textit{CM\_G} and \textit{CM\_G\_TG}), have considerably less process migrations and context-switches than \textit{NONE} and \textit{CM} baseline scenarios. Moreover, for \textit{CM\_G*} scenarios, as each container runs a single process, this is essentially a single-level scheduling (i.e. at the cgroup level), which is simpler and allows to exploit processor affinity better, in a similar way to when processes are pinned explicitly, which is an important factor for the performance of CPU-intensive applications.

\begin{figure}[htbp]
    \centering
    \includegraphics[width=0.65\linewidth]{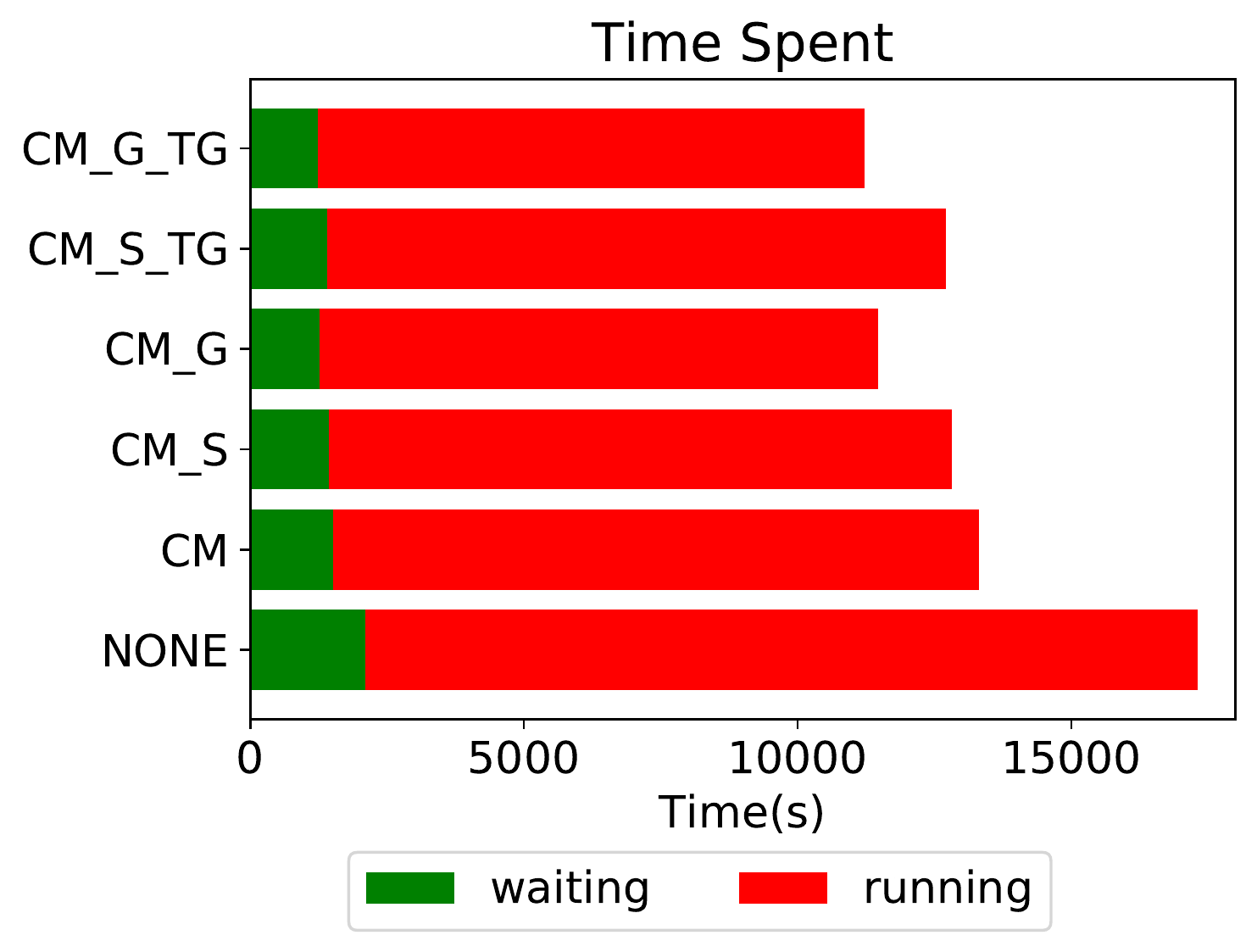}
    \caption{Overall response time of scheduling 10 EP-DGEMM jobs.}
    \label{fig:exp1-makespan}
\end{figure}

\begin{figure*}[hbtp]
    \centering
     \begin{subfigure} 
        \centering 
        \includegraphics[width=0.28\linewidth]{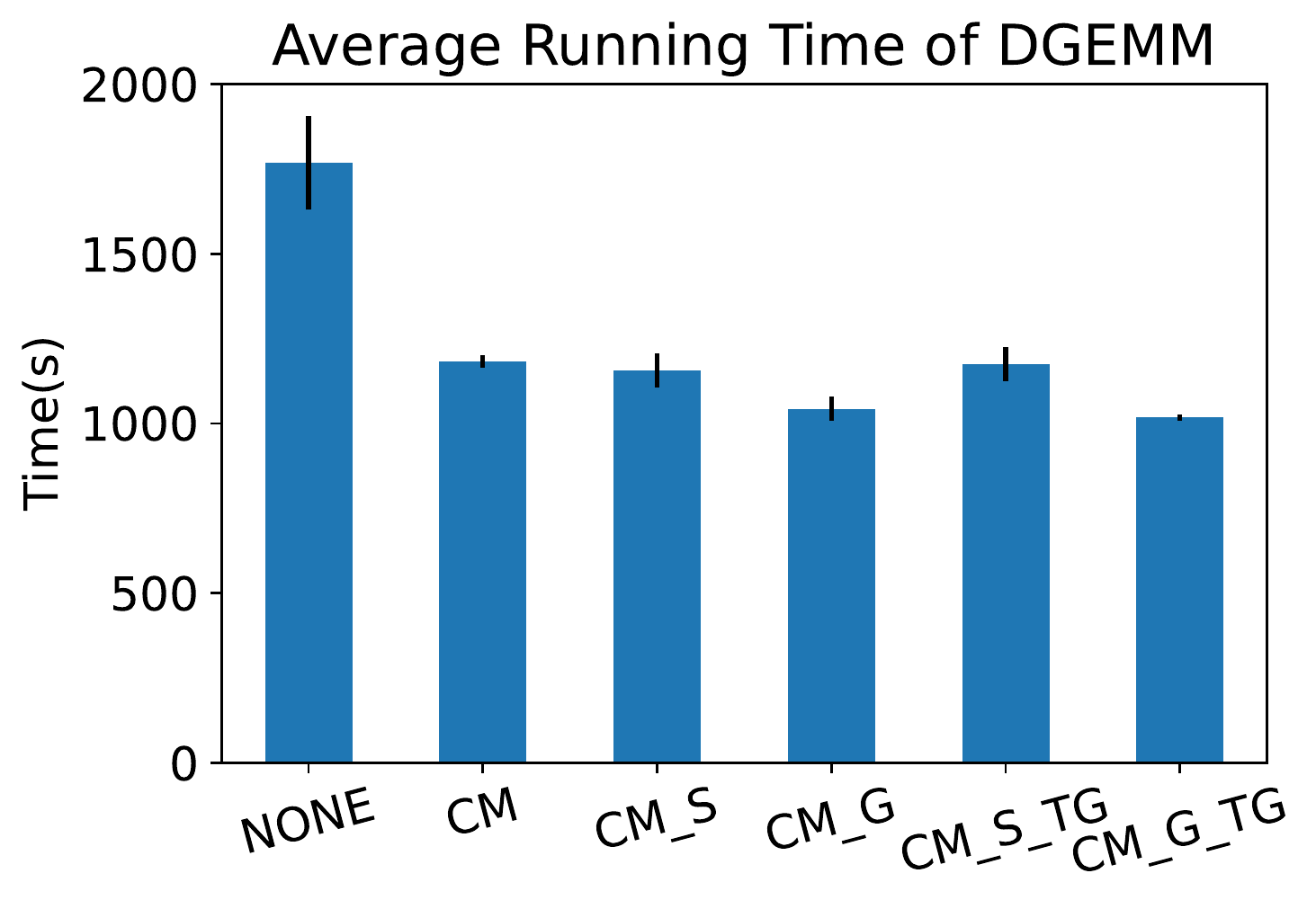}
    \end{subfigure}
    \begin{subfigure}   
        \centering 
        \includegraphics[width=0.28\linewidth]{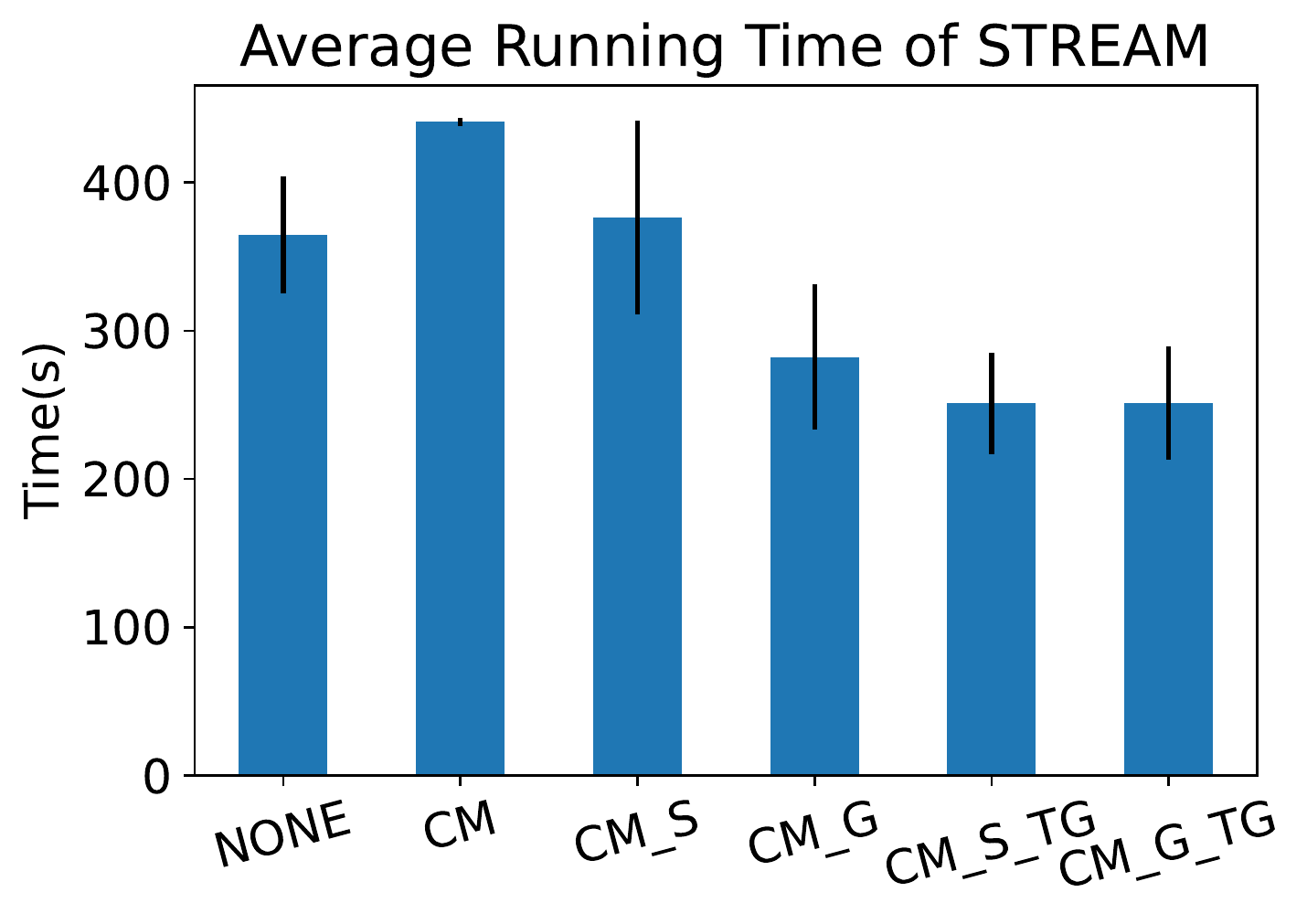}
    \end{subfigure}
    \begin{subfigure}   
        \centering 
        \includegraphics[width=0.28\linewidth]{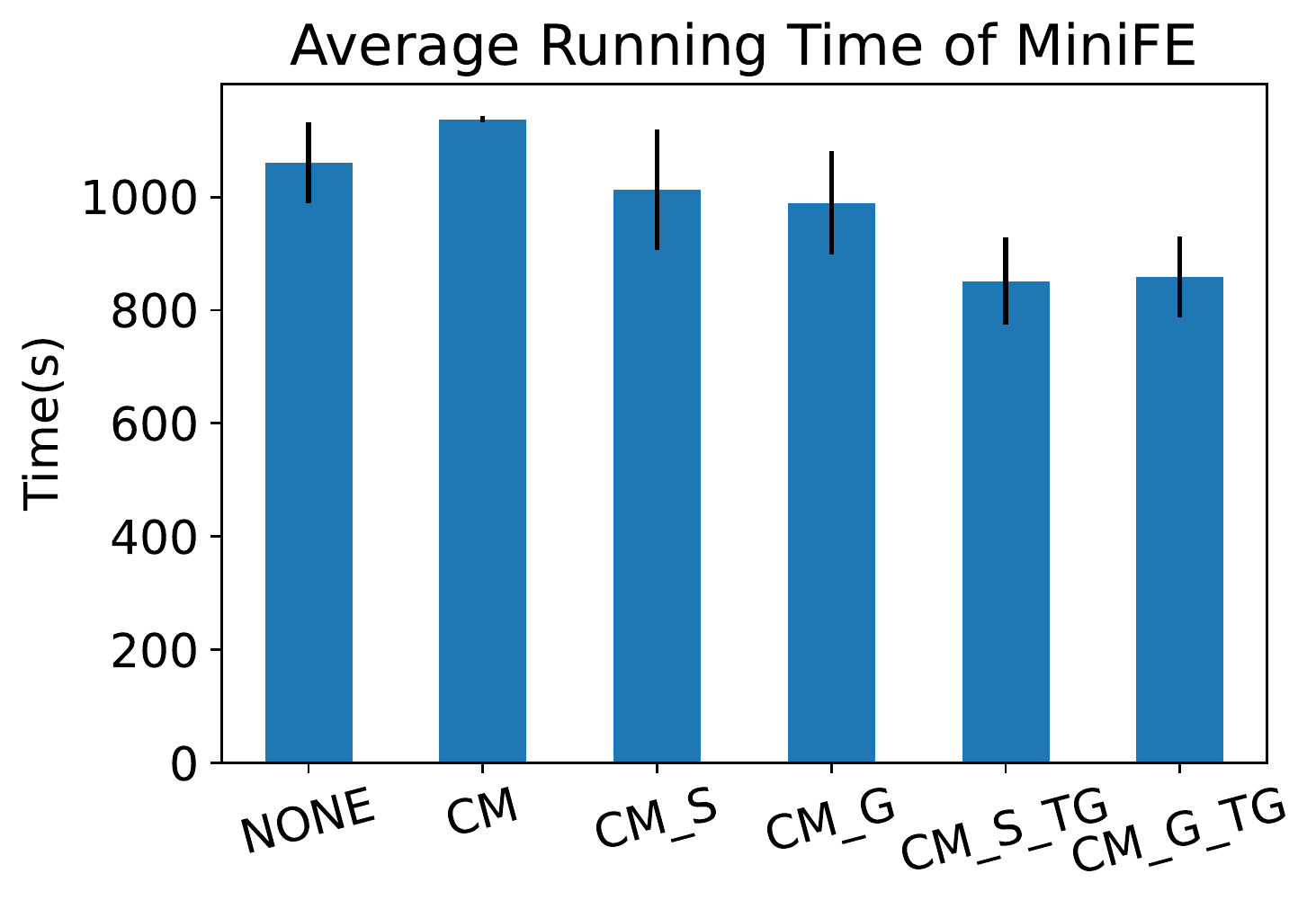}
    \end{subfigure}
    \begin{subfigure}   
        \centering 
        \includegraphics[width=0.28\linewidth]{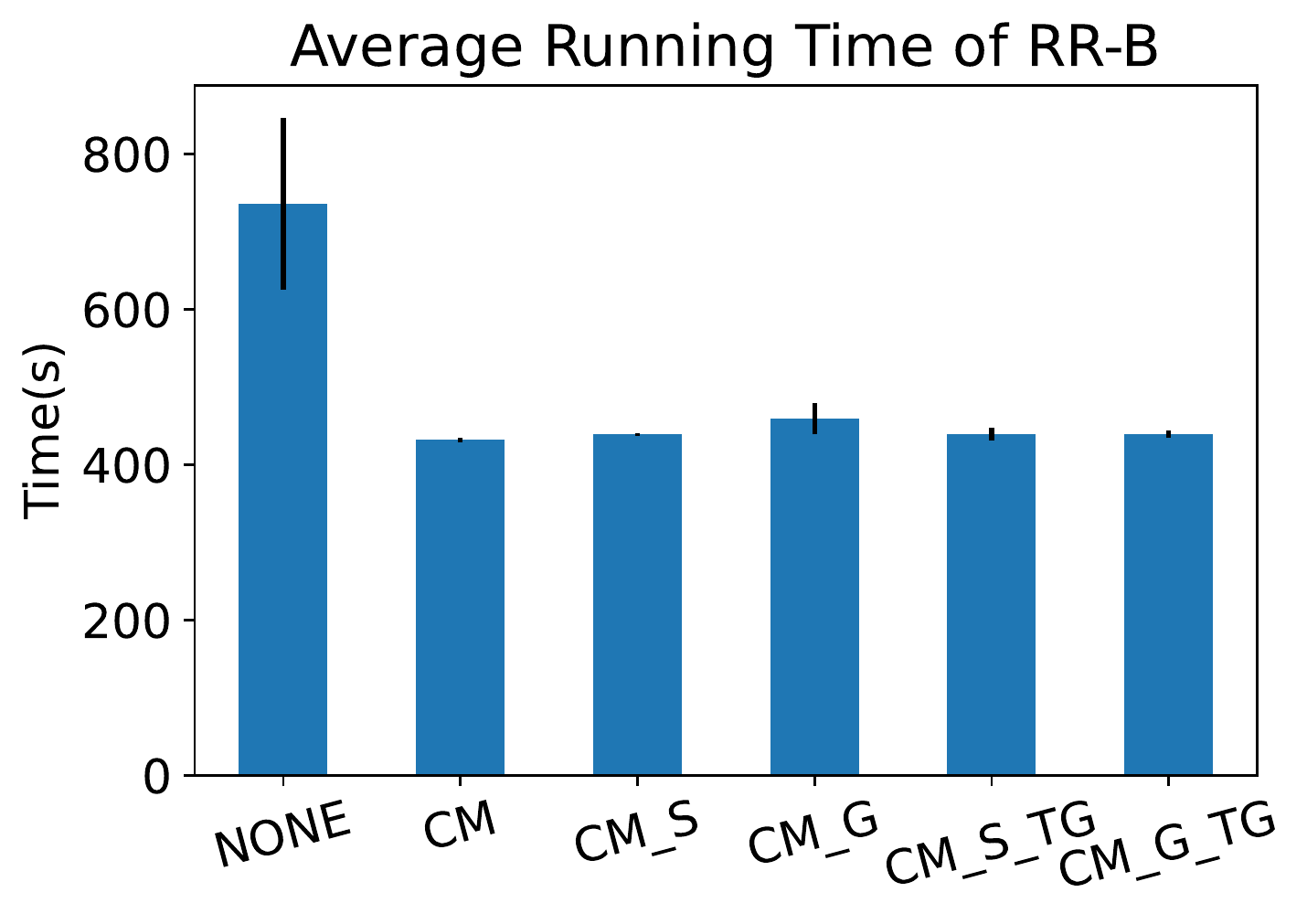}
    \end{subfigure}
    \begin{subfigure}   
        \centering 
        \includegraphics[width=0.28\linewidth]{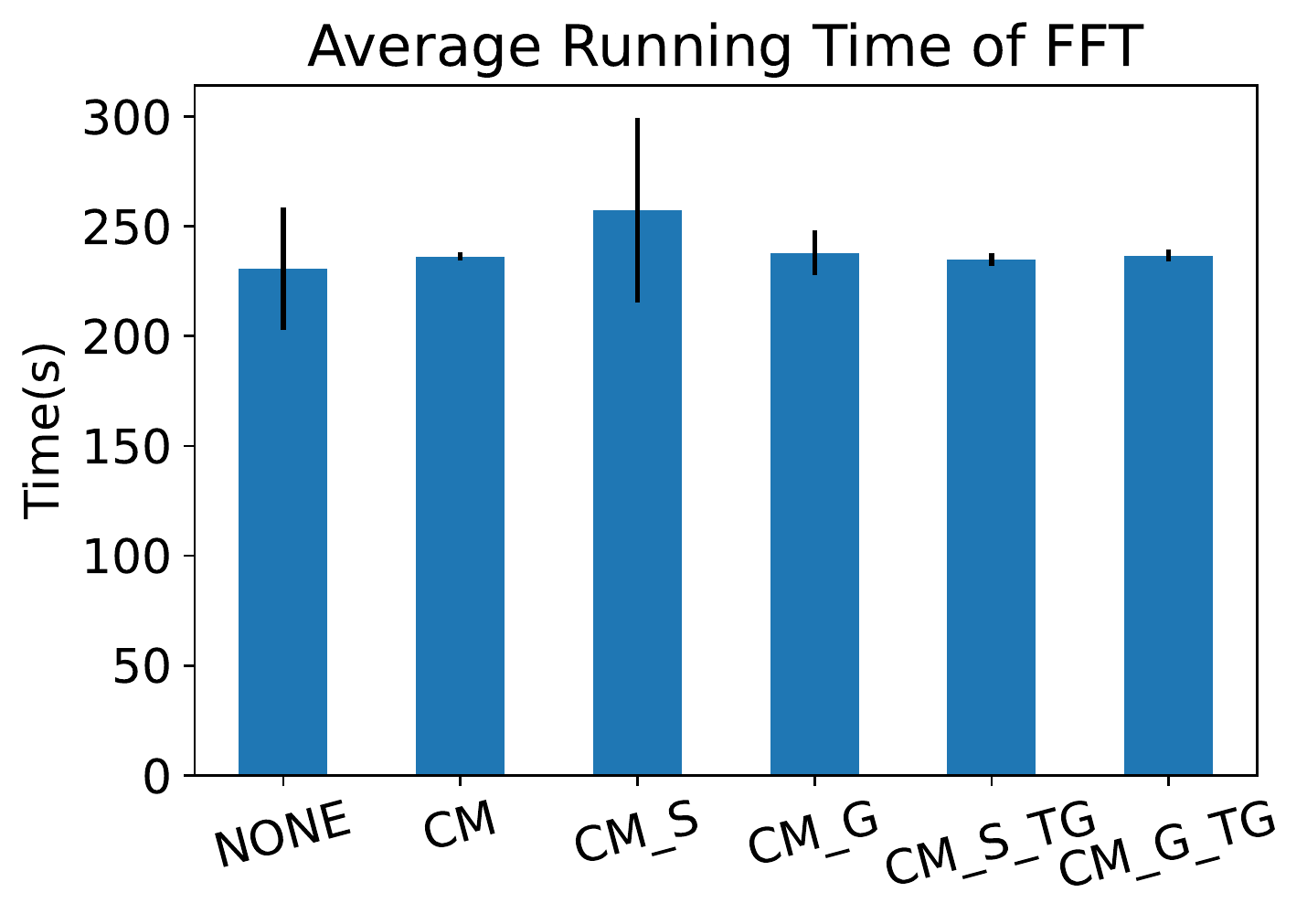}
    \end{subfigure}
    \begin{subfigure}   
        \centering 
        \includegraphics[width=0.28\linewidth]{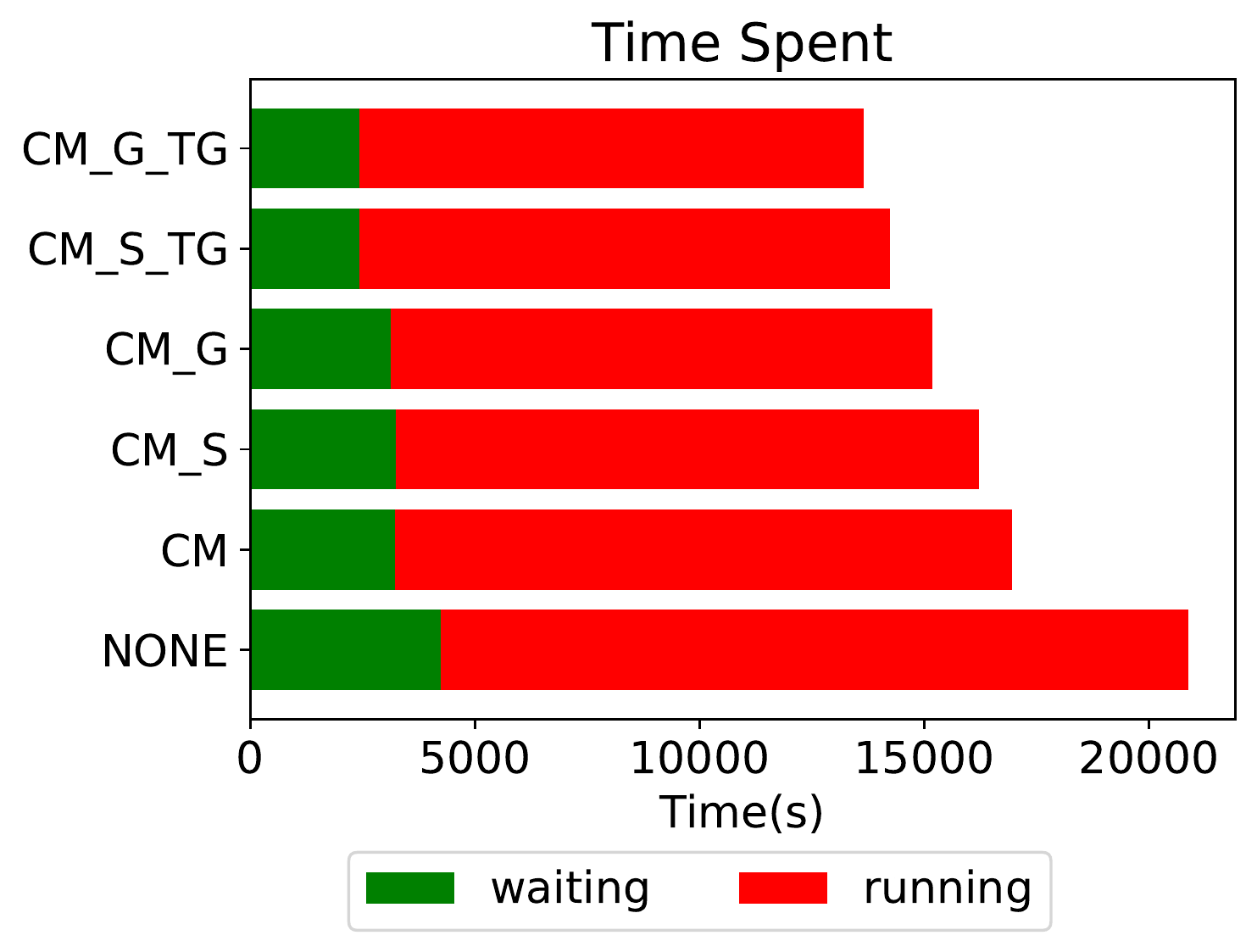}
    \end{subfigure}
    \caption{Results of experiment 2: the first five graphs show the average job running time of each type of workload; the last graph presents the overall response time when scheduling 20 jobs of different types.}
    \label{fig:exp2-time}
\end{figure*}

As shown in Fig. \ref{fig:exp1-makespan}, the improvements in the running time of DGEMM in those scenarios cause also an improvement in the overall response time. In particular, \textit{CM\_S*} have 5\% and 26\% improvement and \textit{CM\_G*} have 15\% and 34\% improvement, compared to baseline scenarios \textit{CM} and \textit{NONE}, respectively. Note that \textit{TG} incurs no significant benefit for DGEMM because its CPU requirements can be granted in all cases and thereby it does not suffer imbalance problems.

\subsection{Experiment 2: Schedule with multiple types of MPI workloads}

In this experiment, we evaluate the effectiveness of our policies to fit different types of workloads. We randomly generate a submission time for 20 MPI workloads within the interval from 0 to 1200 seconds. Workloads come from the five benchmarks (i.e., EP-DGEMM, EP-STREAM, G-FFT, G-RandomRing Bandwidth, and MiniFE), and each benchmark will be run 4 times, with a random sequence.

Fig. \ref{fig:exp2-time} shows the average job running time of several workloads and the overall response time in the six scenarios. Baseline scenario \textit{NONE} uses shared resources for all the running workloads, thus potentially having computation imbalance as the processes can move among the several CPUs in the node. The randomness of these processes movement can incur a variable performance between different executions of the same type of job, thus impacting the average runtime. Baseline scenario \textit{CM} shows better cache utilization (less L3 misses) and reduces remote memory accesses latency, but introduces more memory contention for memory-intensive applications than \textit{NONE} scenario. 

When enabling 'scale' and 'granularity' policies in \textit{CM\_S*} and \textit{CM\_G*}, we partition CPU- and memory-intensive applications with more number of containers but less number of processes per container, while the processes within a network-intensive application remain in a single container to avoid the network latency. As shown in Fig. \ref{fig:exp2-time}, 'scale' and 'granularity' policies do not have significant effect on the network-intensive applications (i.e., RR-B and FFT), but improve considerably the performance of CPU- and memory-intensive applications regarding the baseline scenarios. 
Task-group scheduling (TG) has an important impact on memory-intensive benchmarks, for instance, \textit{CM\_S\_TG} can reduce a 33\% the running time of STREAM in relation to \textit{CM\_S}. This is because by default the scheduler randomly chooses the nodes to deploy the pods within a same job, and some load imbalance could introduce more memory contention and latency. TG uses even distribution for jobs to deploy their pods into nodes, thus maximally guaranteeing the balance of MPI applications.

Fig. \ref{fig:exp2-time} shows the overall response time of \textit{CM\_S\_TG} has 16\% and 32\% improvement, and \textit{CM\_G\_TG} has 19\% and 35\% improvement, both compared to baseline scenarios \textit{CM} and \textit{NONE}, respectively. These come both from the granularity selection, but also from the task-group scheduling, since \textit{CM\_S\_TG} and \textit{CM\_G\_TG} have 12\% and 10\% performance improvements with respect to \textit{CM\_S} and \textit{CM\_G}. 

\begin{figure*}[hbtp]
    \centering
     \begin{subfigure} 
        \centering 
        \includegraphics[width=0.28\linewidth]{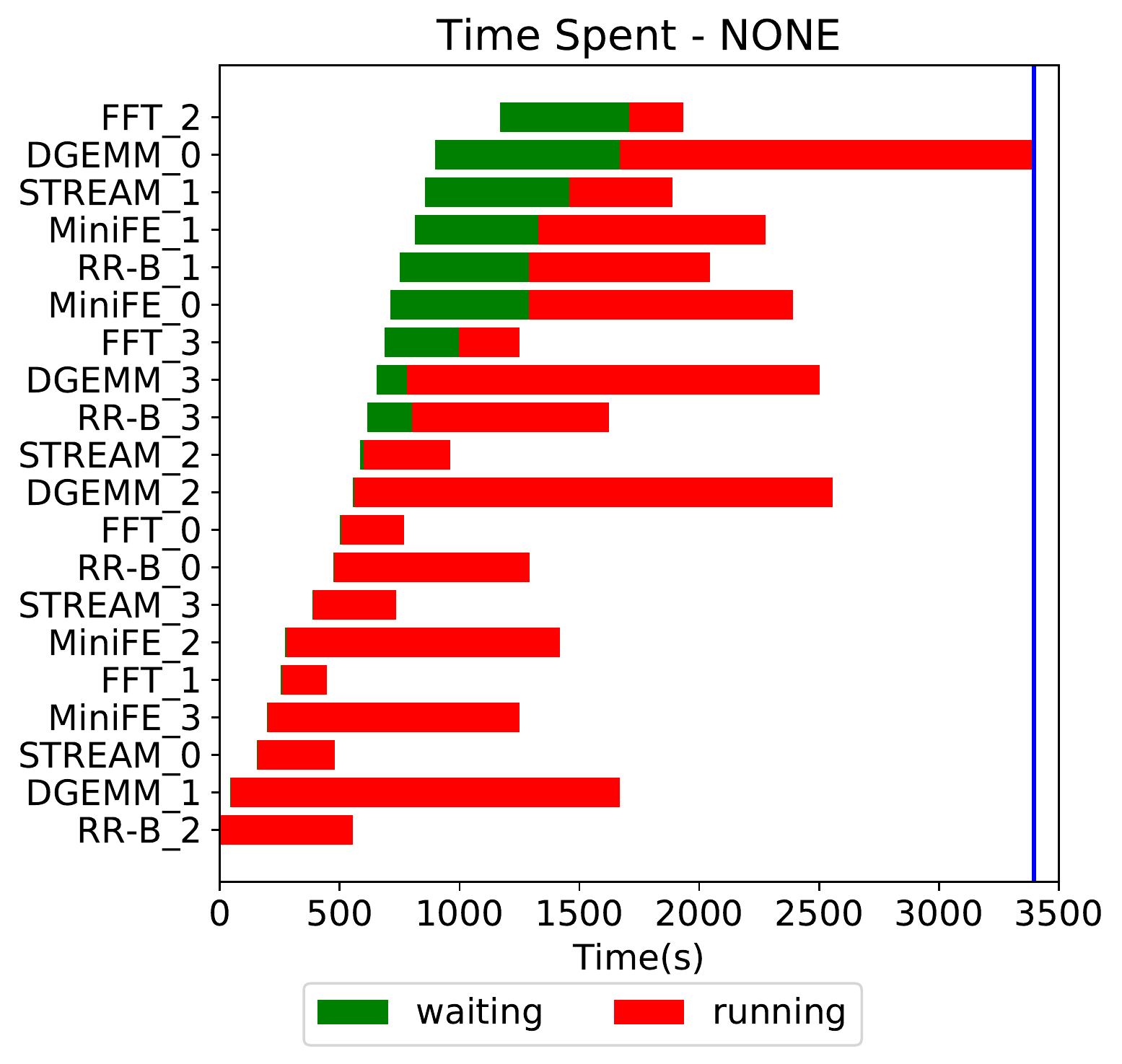}
    \end{subfigure}
    \begin{subfigure}   
        \centering 
        \includegraphics[width=0.28\linewidth]{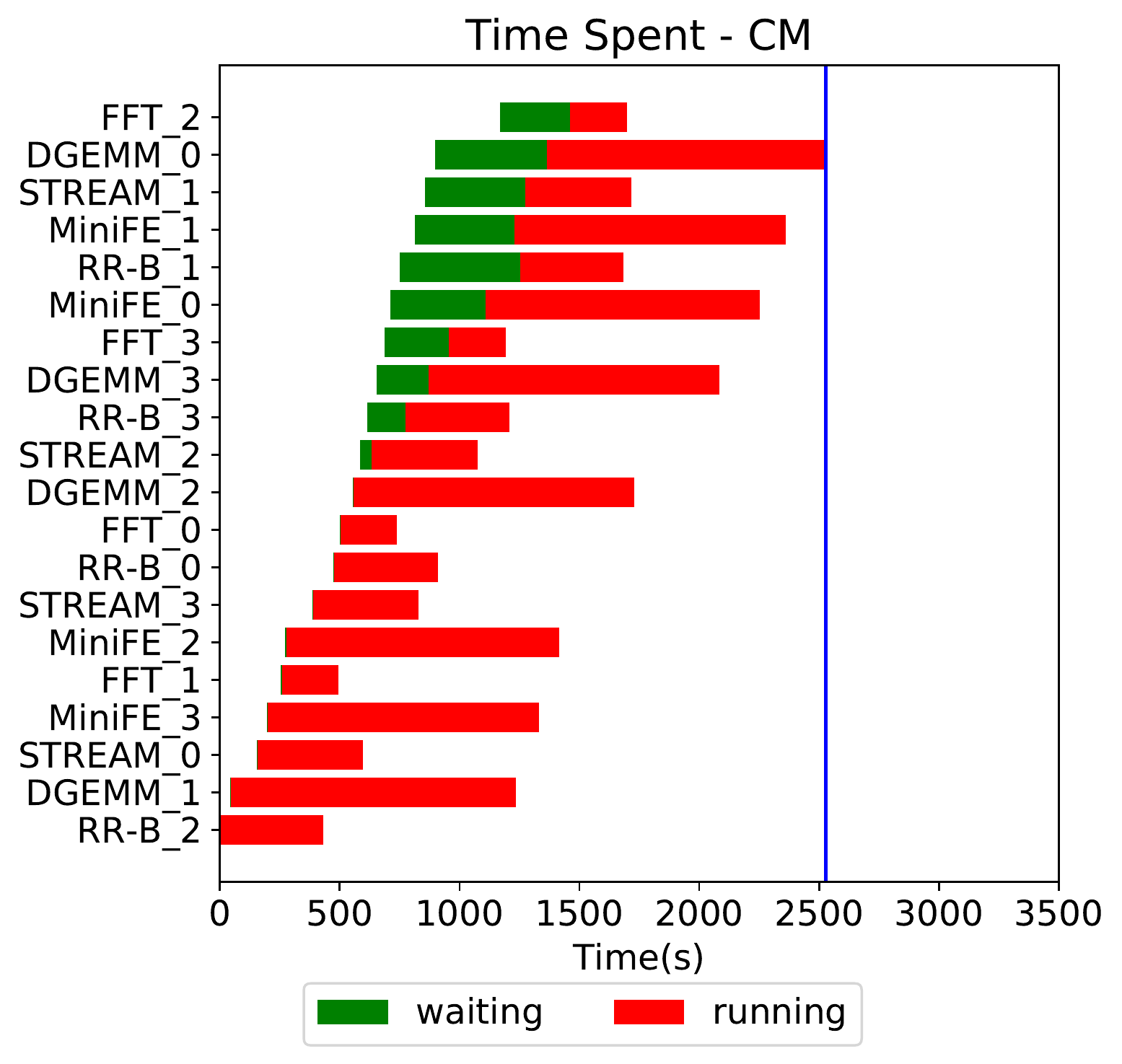}
    \end{subfigure}
    \begin{subfigure}   
        \centering 
        \includegraphics[width=0.28\linewidth]{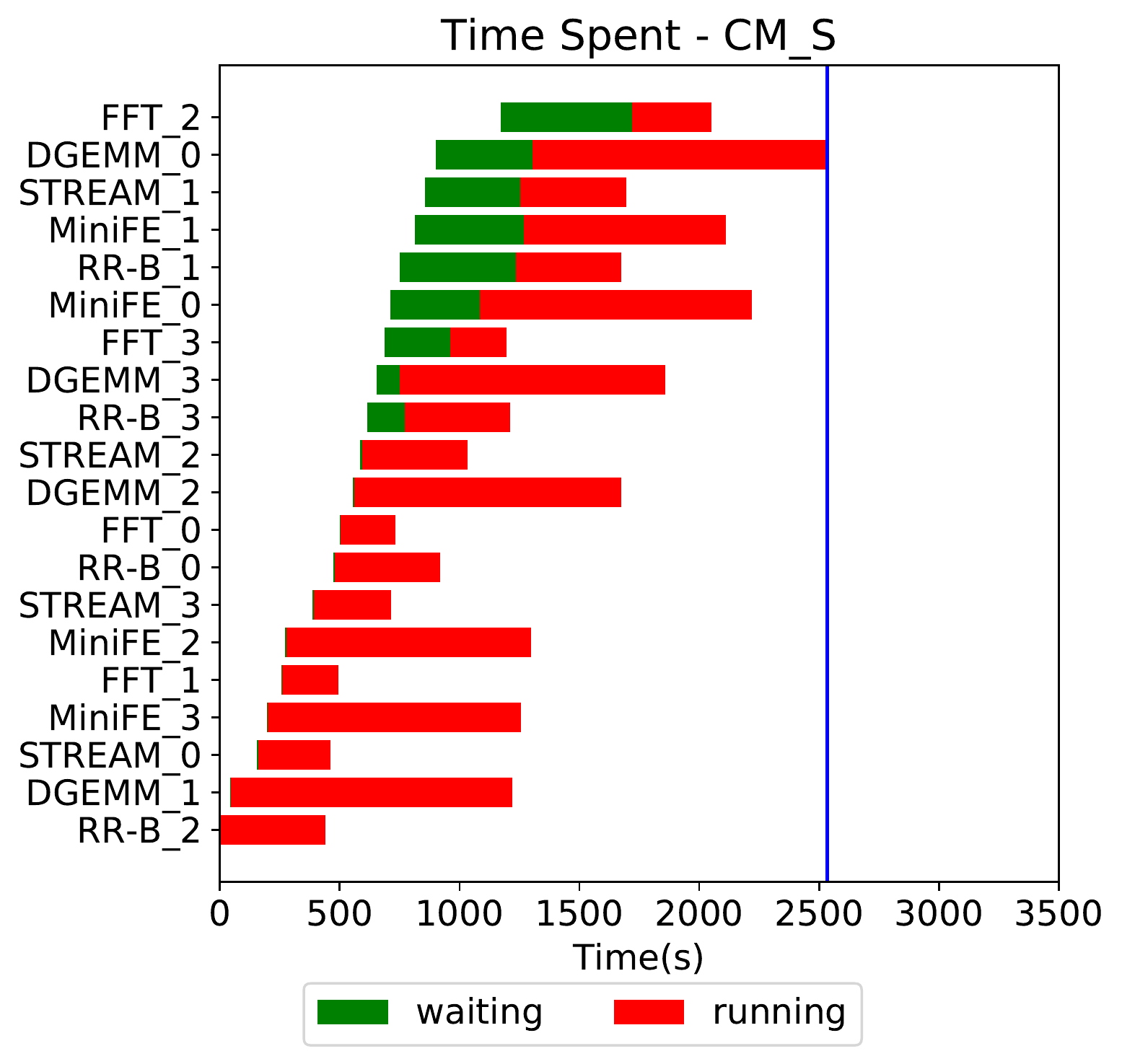}
    \end{subfigure}
    \begin{subfigure}   
        \centering 
        \includegraphics[width=0.28\linewidth]{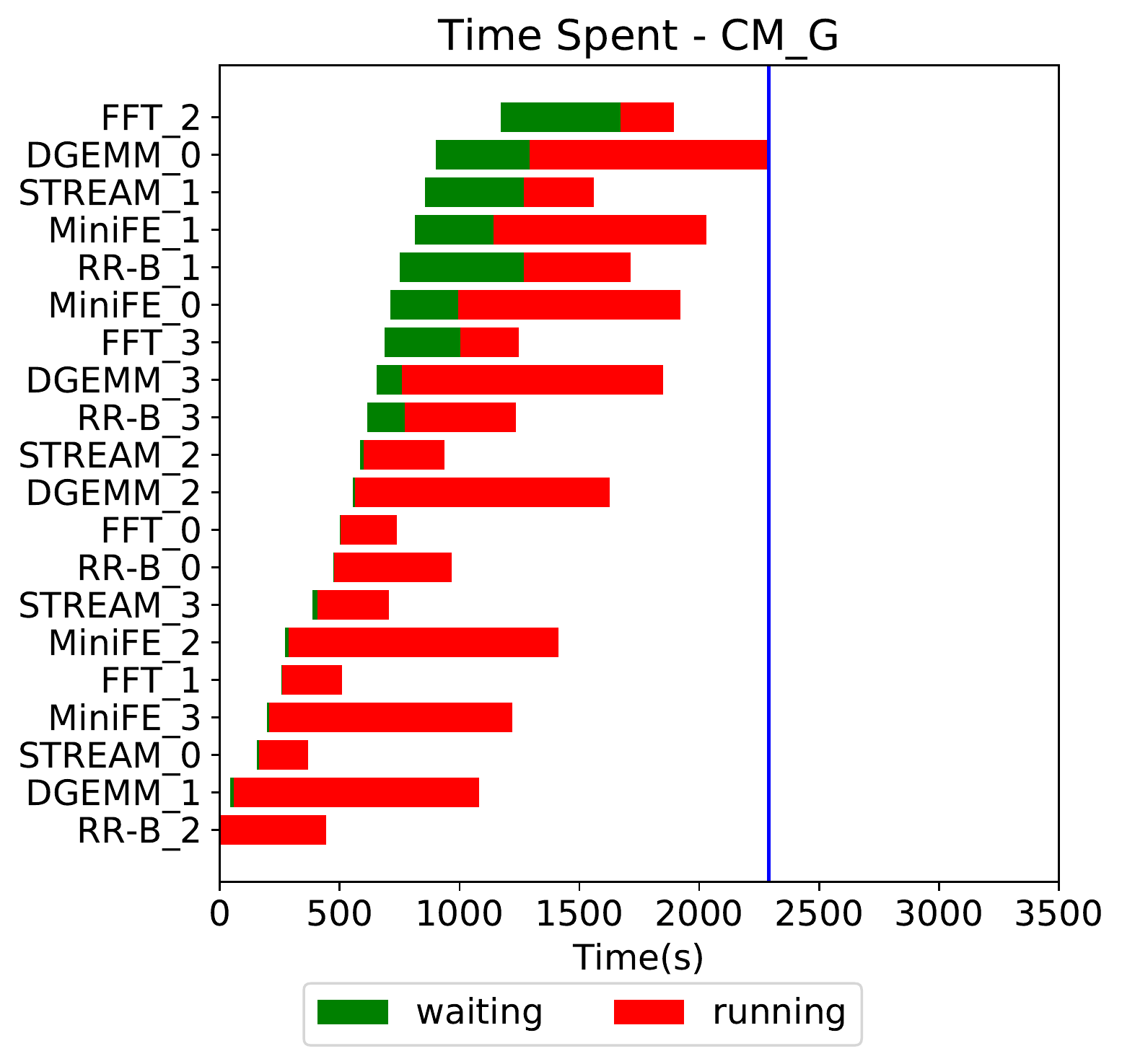}
    \end{subfigure}
    \begin{subfigure}   
        \centering 
        \includegraphics[width=0.28\linewidth]{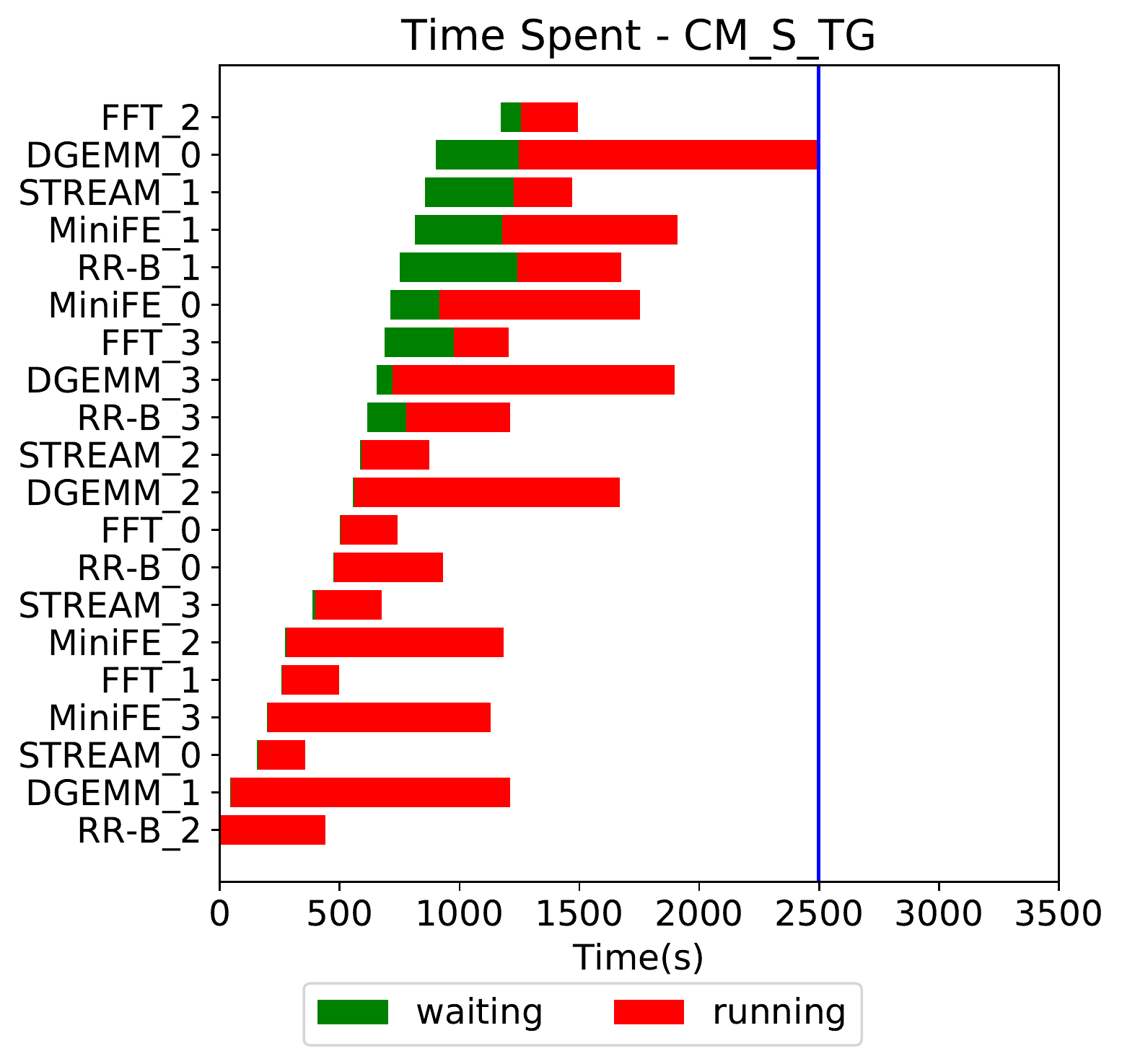}
    \end{subfigure}
    \begin{subfigure}
        \centering 
        \includegraphics[width=0.28\linewidth]{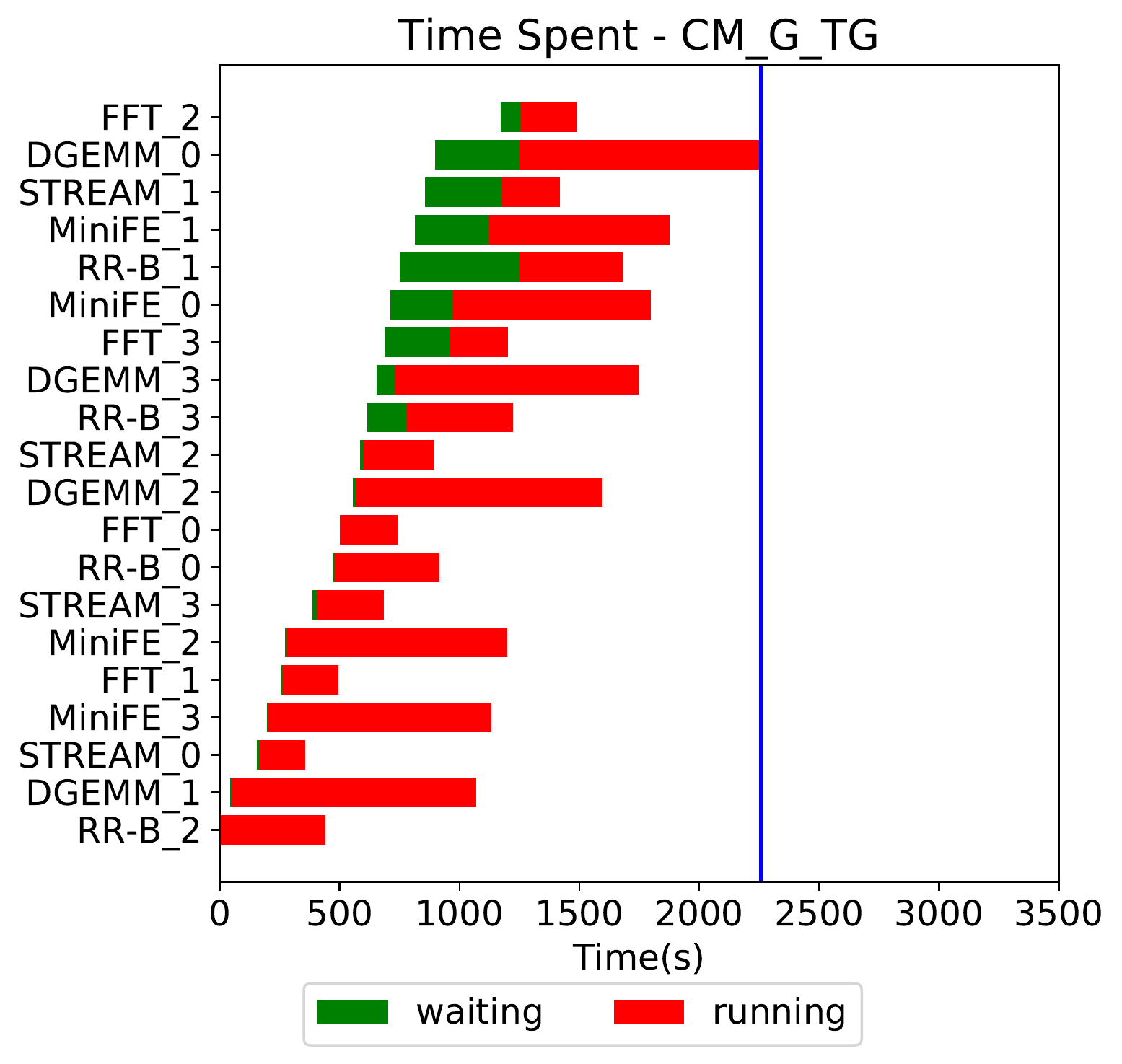}
    \end{subfigure}
    \caption{Makespan of experiment 2: scheduling 20 jobs of different types.}
    \label{exp1-result}
    \vspace{-0.8em} 
\end{figure*}

To evaluate the effectiveness of our two-level scheduler for the entire workload, we show the makespan in Fig. \ref{exp1-result}, 
which also presents in detail the scheduling process of each scenario. Scenario \textit{CM\_S\_TG} has 1\% and 26\% makespan reduction, whereas scenario \textit{CM\_G\_TG} has 11\% and 34\% makespan reduction, both with respect to baseline scenarios \textit{CM} and \textit{NONE}, respectively, which demonstrate how our policies could improve overall system throughput. 



\begin{figure}
    \centering
    \includegraphics[width=\linewidth]{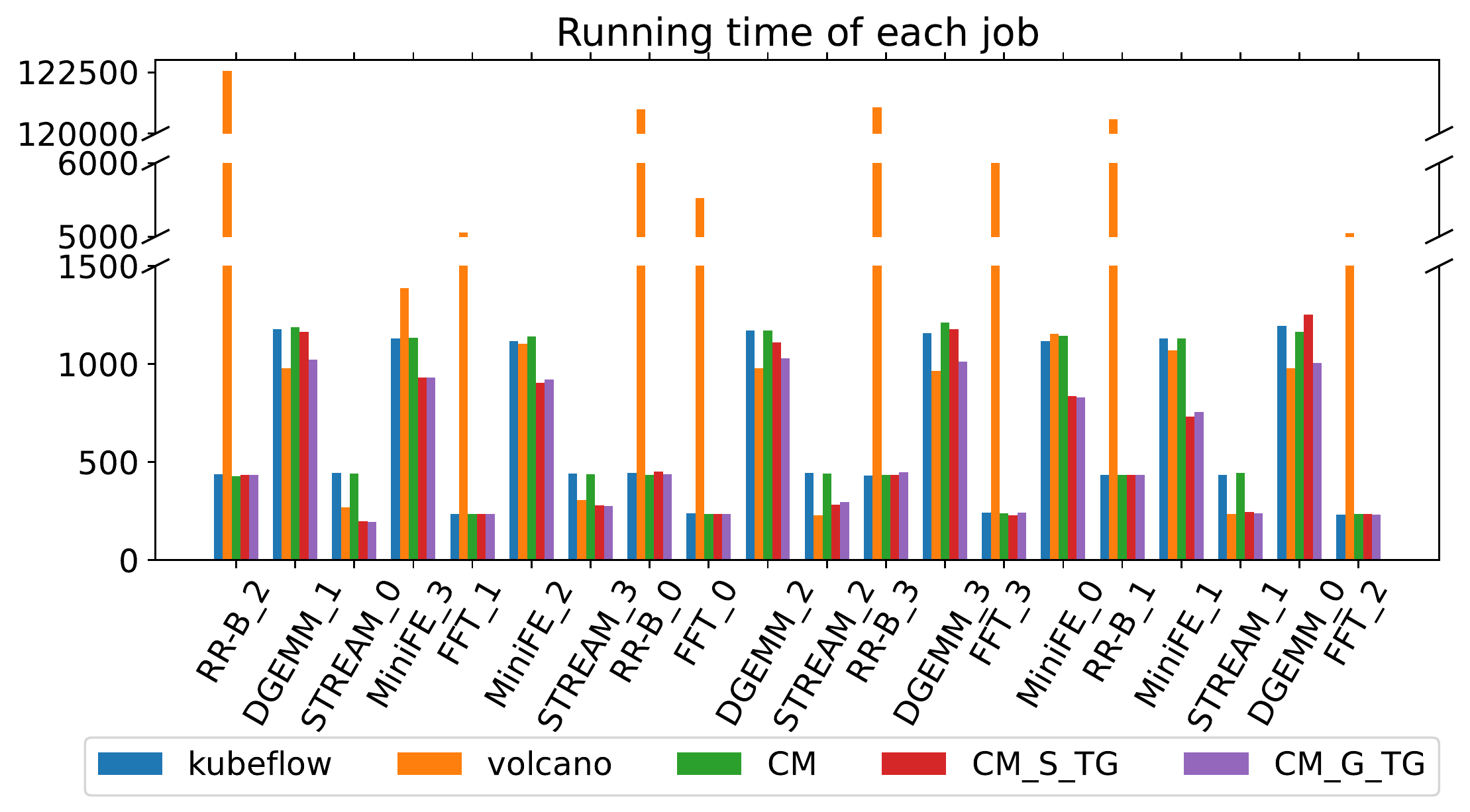}
    \caption{Job running time with different frameworks.}
    \label{fig:exp5-jobrun}
\end{figure}
\begin{figure}
    \centering
    \includegraphics[width=\linewidth]{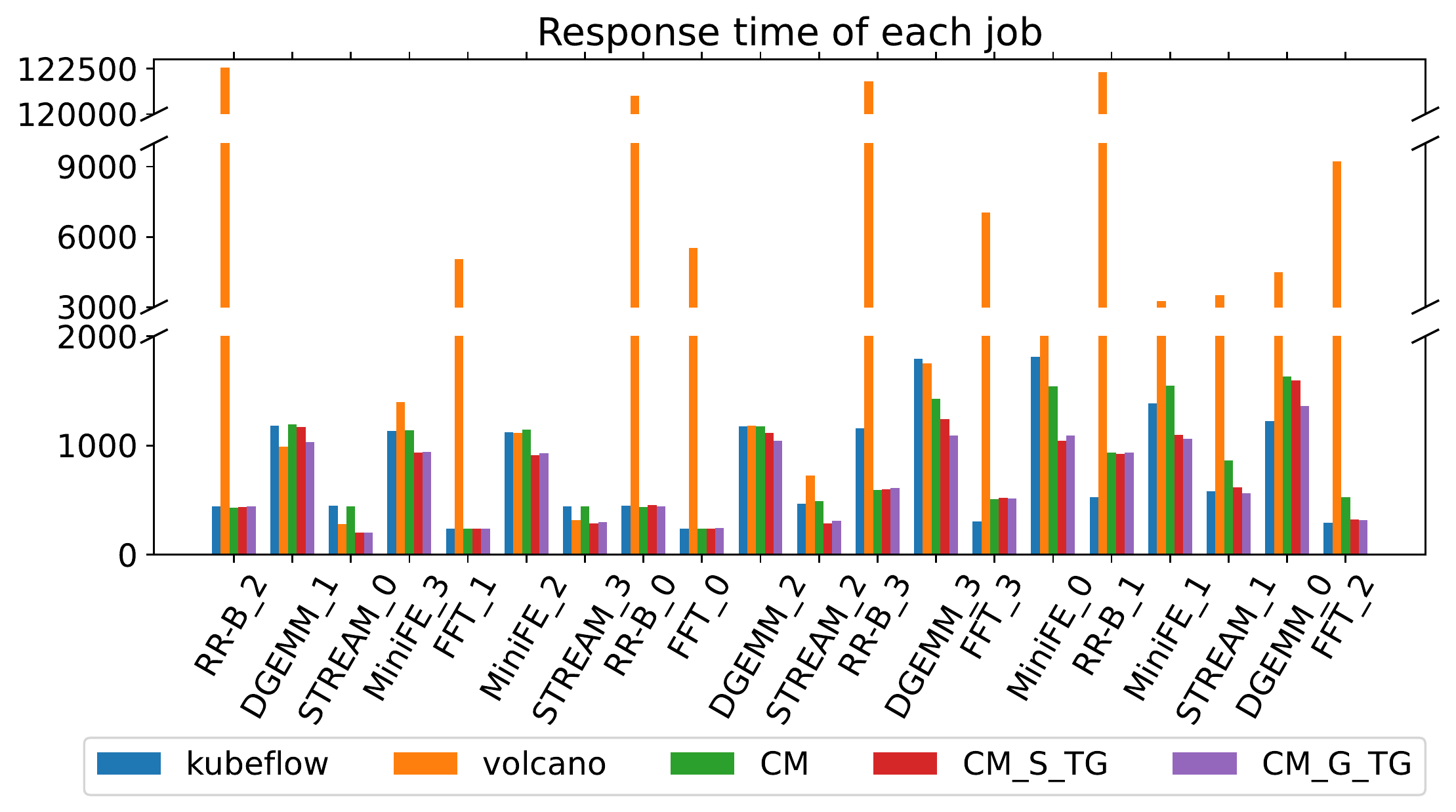}
    \caption{Job response time with different frameworks.}
    \label{fig:exp5-jobcomplete}
\end{figure}

\subsection{Experiment 3: Schedule Under Different Frameworks}
This experiment compares our approach to schedule MPI workloads with Kubeflow MPI operator\cite{kubeflowmpi} and native Volcano\cite{volcanompi}. MPI jobs specified by Kubeflow are scheduled by Kubernetes default scheduler. Volcano specifies jobs through its own Job Controller and schedules them using Volcano Scheduler, which features a gang plugin by default. Kubelet for these two scenarios is set with CPU/memory affinity enabled. Other settings are the same as experiment 2.

As shown in Table \ref{tab:makespan}, which displays the makespan for single executions under all the evaluated scenarios, \textit{Kubeflow} framework has similar makespan to the \textit{CM} baseline scenario, because both use CPU/memory affinity settings and use the default or default-alike scheduler. \textit{Volcano} framework has an important slowdown on makespan, because it partitions all the workloads, even the network-intensive ones, which incur high latency and contention. Consequently, both frameworks fail to provide better performance than our fine-grained scheduling.

\begin{table}[htbp]
\caption{Makespan comparison. }
\centering
\begin{tabular}{p{2.8cm}|p{4cm}}
\hline
\textbf{Scenarios} & \textbf{Makespan}  \\ \hline
Kubeflow &  0 days, 00:42:00 (2520 s)\\ 
Volcano  &  1 days, 10:10:55 (123055 s)\\ 
CM   &  0 days, 00:42:09 (2529 s)   \\
CM\_S\_TG  &   0 days, 00:41:38 (2498 s)      \\ 
CM\_G\_TG  & 0 days, 00:37:38 (2258 s)    \\ 
\hline
\end{tabular}
\label{tab:makespan}
\end{table}

Fig. \ref{fig:exp5-jobrun} shows the job running time of each job. \textit{Kubeflow} has a similar job running time as \textit{CM}, because they do not partition a job into multiple containers, hence CPU- and memory-intensive workloads cannot benefit from multi-container deployments. Contrariwise, \textit{Volcano} allocates a job by default as one process per container, and those containers are randomly submitted to multiple nodes. Consequently, network-intensive workloads face very important performance degradation due to an increasing number of communications. In scenarios \textit{CM\_S\_TG} and \textit{CM\_G\_TG}, some of the CPU- and memory-intensive workloads show even better performance than \textit{Volcano} because those scenarios enable the task-group plugin so that the group of fine-grained containers from a same job can be evenly allocated to the nodes.

Fig. \ref{fig:exp5-jobcomplete} shows the job response time of each job. Our fine-grained scheduling outperforms the rest, in particular, the container allocation in \textit{CM\_G\_TG} scenario improves (or at least equals) the running time of all the jobs, as well as their waiting time. \textit{Volcano} is the worst case, as network-intensive workloads have an important performance degradation, thus also introducing more waiting time for the following jobs.

%% file: 7.conclusion.tex
\section{Conclusion}\label{sec:conclusion}
This paper presented fine-grained scheduling policies for allocating containerized HPC workloads in a Kubernetes cluster. We extended the Scanflow-Kubernetes platform to support HPC MPI workloads and improved its two-layer scheduling architecture, by creating new policies in both the application-layer planner agent (i.e., enabling granularity selection), as well as the infrastructure-layer Volcano controller and scheduler (i.e., adding an MPI-aware controller and a task-group scheduling plugin). 

Our results show that the proposed fine-grained policies can reduce the response time of HPC workloads up to 35\%, as well as improve the makespan up to 34\%. Although our benchmarks are small-scaled MPI jobs that fit in a single node, our principles to exploit granularity are also applicable if applications do not fit in a single node: e.g. for network applications, one would probably use coarse-grained granularity within each node to exploit fast shared-memory communication, whereas CPU-bound applications could use fine-grained granularity to exploit affinity.
In the future, we will enhance our fine-grained policies for the scheduling of mixed HPC-AI workloads on Kubernetes, and to consider other application profiles such as I/O applications. Moreover, we will evaluate them in larger-scale scenarios.